\documentclass[two-columns]{nature} 
\usepackage{graphicx,caption} \usepackage{color} \usepackage{soul}
\usepackage{siunitx} \usepackage{amsmath,physics}
\usepackage[unicode=true,bookmarks=true,colorlinks=true,
allcolors=blue]{hyperref} \usepackage{cleveref}
\graphicspath{{./images/}} \usepackage{todonotes}
\usepackage{textcomp} 
\usepackage{gensymb}

\usepackage{lineno}

\newcommand{\mb}[1]{\mathbf{#1}}

\usepackage[backend=biber,bibencoding=utf8, style=nature, url=false, isbn=false, doi=false]{biblatex}
\title{Broadband vectorial ultra-flat optics with experimental efficiency up to 99\% in the visible via universal approximators}

\author{F. Getman$^{1\dag}$, M. Makarenko$^{1\dag}$, A. Burguete-Lopez$^{1\dag}$
\& A. Fratalocchi$^1$}

\begin{document}

\maketitle

\begin{affiliations}

\item PRIMALIGHT, Faculty of Electrical Engineering; Applied Mathematics and
    Computational Science, King Abdullah University of Science and Technology,
    Thuwal 23955-6900, Saudi Arabia 

$^\dag$ These authors contributed equally to this work.  \end{affiliations}
\begin{abstract} 
Integrating conventional optics into compact nanostructured surfaces is the goal of flat optics. 
Despite the enormous progress of this technology, there are still critical challenges for real world applications due to a limited efficiency in the visible, on average lower than $60\%$, which originates by absorption losses in wavelength thick ($\approx 500$~nm) structures. Another issue is the realization of on-demand optical components for controlling vectorial light at visible frequencies simultaneously in both reflection and transmission, and with a predetermined wavefront shape. 
 In this work, we developed an inverse
    design approach that allows the realization of highly efficient (up to $99\%$) 
    ultra-flat (down to $50$~nm thick) optics for vectorial light control and broadband input-output responses on a desired wavefront shape. The approach leverages on a hidden network of universal approximators, which exist in the physical layer of suitably engineered semiconductor nanostructures. Near unity performance results from the ultra-flat nature of these surfaces, which reduces absorption losses to almost negligible values. Experimentally,
    we discuss polarizing beam splitters, comparing their performances with the best results obtained from both direct and inverse design techniques, and new flat-optics components represented by dichroic mirrors and the basic unit of a flat optics display 
    that creates full colors by using only two
    sub-pixels, overcoming the limitations of
    conventional LCD/OLED technologies that require three sub-pixels for each composite color. Our devices are manufactured with a
    complementary metal-oxide-semiconductor (CMOS) compatible
     process, making them scalable for mass
    production at inexpensive costs.  
    \end{abstract}

\section*{Introduction}


The production of lightweight and wearable opto-electronic devices is hampered by the bulk and expensive nature of
traditional optical components~\cite{Yu2014,Chen2018,He2018}. Flat optics aims to address this problem by replacing conventional optics with highly integrated nanostructured surfaces~\cite{Zhou2017,Guo2018,He2018,Devlin2016,Shrestha2018,Glybovski2016, Colburn2014}.
This technology has attracted an enormous interest, demonstrating a large variety of designs ranging from
lenses~\cite{Khorasaninejad2016,Zheng2017,Chen2018,Shrestha2018,Wang2018}, to
holograms~\cite{Arbabi2015,Chong2016,Wang2016a, Ren2019, Huang2018},  filters~\cite{Wood2017,doi:10.1002/adom.201900849,Galinski2017,Arbabi2015,Guo2017,Li2019}, and to other  components that outperform their traditional optical counterparts~\cite{Decker2016, Staude2017, Chang2018, Yang2017, Kuznetsov2016, Jahani2016, Colburn2014,Glybovski2016,Khorasaninejad2016,Slovick2017a, Overvig2019}. Currently, challenges that are being addressed at visible frequencies are related to the scalability of the structure fabrication, the design of different types of broadband functionalities, and the increase of efficiency~\cite{Liu2018a,Chen2019, Shibanuma2016}.\\  
The majority of flat optics components designed to operate in the visible exploit propagation phase shifts in truncated waveguides~\cite{Colburn2014}. 
This approach requires wavelength thick structures, ranging between \SI{235}{nm} to
\SI{800}{\nano\meter}, 
with absorption limited efficiencies between 20\% and
67\% for designed elements such as polarization splitters~\cite{Li2019,Guo2017}.
While materials such as $\text{TiO}_2$ showed promising results with this scheme, reaching efficiencies up to 86\% for a single frequency metalens at $405$~nm,
 they require at the core of the manufacturing the use of atomic layer deposition (ALD), a process that is not scalable for mass production due to deposition rates of less than one angstrom per cycle~\cite{Khorasaninejad2016, Chen2019,Devlin2016,BalthasarMueller2017,Colburn2014,Johnson2014}. \\ 
Another class of flat optics devices recently developed exploits the interaction between suitably engineered electric and magnetic dipoles, reporting deflectors
and  lenses
 with efficiencies between 58\% and 78\% \cite{Ollanik2018a,Cheng2016,Aoni2019,Yu2015}. Electric and magnetic dipoles, however, allow to control the system response within the bandwidth in which the dipoles overlap, which is typically a single frequency~\cite{Kirill2019,ToteroGongora2017,Ollanik2018a,10.1093/nsr/nwy017}.\\  
 The method of inverse design has been recently investigated as a alternative route to intuition based flat optics designs~\cite{SeanMolesky2018}. In this approach, the desired response is set as the input parameter and the computer furnishes the material design by machine learning. Currently, approaches being explored are centered on the optimization of structures with periodic cells of assumed known period~\cite{Malkiel2018,Lin2020,Nadell2019,Liu2018b}, or on the generation of complex patterns from random distribution of refractive indices~\cite{Liu2018a,Sell2017a,Sell2017b,Phan2019,Andkjaer2014}. The best results available today at near-infrared~\cite{Shen2014,Sell2017b,Sell2017a,Singh2019} and gigahertz~\cite{Callewaert2018} are up to $77\%$ efficiency for polarizer beam splitters, while in the visible they reach up to 67\% for metalenses~\cite{Phan2019}, reporting performances still comparable to the best available designs based on intuition.\\
 A present limitation of inverse design is the lack of a universal strategy that is guaranteed to produce working devices with high efficiency: it is well known that optimization theory fails if the initial design is either too far from the solution, or it is developed along directions that are not converging~\cite{Gendreau2019}.\\
 
 A second important question is that all design approaches (direct and inverse) proposed so far typically control the system response either in transmission or in reflection, and do not yet tackle broadband vectorial light management at visible frequencies simultaneously in both reflection/transmission and with a desired wavefront, such as in, e.g., dichroic mirrors, and other standard components that are not yet realized in flat-optics.
 These points are particularly interesting with reference to recent progress in diffractive deep optical networks at THz frequencies~\cite{Lin1004} and in neuromorphic computing with nonlinear waves~\cite{1912.07044,frataxai0}. If it could be possible to integrate some of these universal concepts into a flat optical structure, we could engineer efficient, scalable and on-demand broadband optical components for light processing via flat surfaces.\\
The aim of this article is to explore a path to address concurrently the issues of efficiency, bandwidth, functional response and fabrication scalability. We begin by addressing the question of how wide the spectrum of functionalities that can be designed by a flat optical structure is. We demonstrate that by suitably engineering a physical hidden layer of universal approximators, which inherently exist in any optical nanostructure, it is possible to design flat optical surfaces that can represent arbitrarily defined output electromagnetic responses. We rigorously prove that the system possesses the same universal expressivity of a feedforward neural network with a non-polynomial activation function and variable threshold~\cite{haykin2009neural}. We then use this result to develop an inverse design approach along optimization lines that can engineer structures with high efficiency, and that can be manufactured with a technique compatible with mass production.\\
Universal approximators are enabled in a design strategy that controls by geometrical deformations a sufficiently large number of nanoscale resonances, theoretically equal or larger to the number of design points in the frequency response of the flat-optics device. As this approach does not rely on propagation effects, it allows to shrink down the thickness of the structure to ultra-flat values, as low as 50~nm, reducing absorption losses to negligible values at visible wavelengths, including the blue region. This approach allows to relax constraints on the shape as well. We mathematically demonstrate that these universal approximators can be generated in geometries that are as simple as possible, such as e.g., cuboid nanostructures, which can be manufactured by using a CMOS compatible fabrication process, scalable to mass production by nanoimprinting or deep UV~\cite{Guo2007}.\\
Implementing these results numerically requires performing a global optimization in a sufficiently large space of multi-modal nanoresonators via first-principle simulations, allowing each nanostructure to explore all possible deformations, without making any assumption on the system periodicity. We address this issue by developing a parallel software that couples large scale optimization techniques with the latest generation of neural network in computer vision~\cite{frataxai2020}.\\ 
We validate these results by implementing a series of flat-optics devices for different applications, comparing performances with both direct and inverse designed structures, and introducing new structures that are not yet realized with flat-optics and defined over wideband responses, ranging from 400~nm and spanning the entire visible range. In all of these examples, we report experimentally measured efficiencies between $90\%$ and $99\%$ in the visible, in flat optical structures with thicknesses down to~$50$~nm for broadband vectorial light control simultaneously in both reflection and transmission with the desired wavefront shapes at visible wavelengths. These results show that improving the knowledge on light-matter interactions with strongly multi-modal optical nanostructures helps the engineering of highly performing nanomaterials.

\section*{Results}

\subsection{Theory of flat optics via universal approximators.} Figure~\ref{funo} summarizes the setup and the main idea of this approach. An 
electromagnetic wave composed of a spectrum of waves with amplitudes $[s_{i1}(\omega),...,s_{in}(\omega)]=\mb {s}_i(\omega)$ (Fig. \ref{funo}a~red arrow) impinges on an optical surface made by complex distributions of dielectric nanostructures grown on a transparent
substrate (Fig. \ref{funo}b), generating both reflected $s_{-n}(\omega)$ and transmitted
$s_{+n}(\omega)$ contributions propagating on different scattering directions (Fig. \ref{funo}a, orange arrows). The system output
response $\mb s_{o}(\omega)=\left[s_{\pm 1}(\omega),...,s_{\pm n}(\omega)\right]$ is composed by the vector containing all the scattered contributions emanating from the surface. The coefficients $s_{ij}$ and $s_{\pm n}$ represent the scalar amplitude of impinging and scattered waves, respectively, and represented e.g., by planar waves. The main question we aim to answer is whether it is possible to design the surface of Fig. \ref{funo}a to act as a universal approximator of arbitrarily defined input-output transfer functions $\mb H(\omega)=\frac{\mb s_o(\omega)}{\mb s_i(\omega)}$.\\
Figure
\ref{funo}c represents the input-output
relationship of the system, obtained from the generalized scattering theory, which provides an equivalent formulation of Maxwell equations
based on an intuitive division of the space into propagation and resonant
effects~\cite{Totero_Gongora_2017,2002.08121,Kirill2019}. A detailed demonstration of this result is presented in~\cite{2002.08121,Totero_Gongora_2017}, while here we summarize the main aspects. In this representation, 
space is partitioned into two main sets: the resonant
nanostructures (resonances space) and the remaining space composing the
external environment (propagation space).\\
Each nanostructure is seen from the
outer space as an ideal perfect electric conductor (PEC) material possessing no
resonance. The outer space is seen from each resonant nanostructure as a perfect
magnetic conductor (PMC) material. With this representation, nanostructures are described by a set of $m=1,...,M$ resonant modes of ideal PMC cavities, while the outer space is characterized by propagation effects of light scattered by an array of ideal PEC structures. The total output $\mb s_o(\omega)$ from the system, resulting from propagation and resonance effects, reads: 
\begin{equation} \label{cont0} 
\mb s_o(\omega)=\mb C\cdot
    \mb s_i(\omega)-\boldsymbol\beta\cdot\boldsymbol\sigma\left[\omega-(\boldsymbol\Omega+i\boldsymbol\Gamma)\right]\cdot
    \mb s_i(\omega),
    \end{equation}
    where $\boldsymbol\sigma\left[\omega-(\boldsymbol\Omega+i\boldsymbol\Gamma)\right]=\frac{\mb K}{i(\omega-\boldsymbol\Omega)+\boldsymbol\Gamma}$,  $\boldsymbol\Omega$ is a diagonal matrix containing the resonant frequencies $\Omega_m=(\boldsymbol{\Omega})_{mm}$ of the $m=1,...,M$ modes of the resonator space, $\boldsymbol\Gamma=\frac{\mb K\mb K^\dagger}{2}$ is the matrix of damping coefficients, $\mb K$ accounts for the coupling between the resonator space and environment, $\boldsymbol{\beta}=\mb C\cdot\mb K^\dag$, and $\mb C$ is a scattering matrix with $\mb C^\dag\mb C=1$. 
    Equation \eqref{cont0} is composed of two main
    contributions: non-resonant $\mb C\cdot\mb s_i(\omega)$ and resonant $\boldsymbol\beta\cdot\boldsymbol\sigma\left[\omega-(\boldsymbol\Omega+i\boldsymbol\Gamma)\right]\cdot\mb s_i(\omega)$
    effects.\\
    The resonant contribution (Fig. \ref{funo}c, blue blocks), is equivalent to a feed-forward neural network (Fig. \ref{funo}d, network inside dashed red box), which processes the spectral frequency $\omega$ at the input into the output $\boldsymbol\beta\cdot\boldsymbol\sigma\left[\omega-(\boldsymbol\Omega+i\boldsymbol\Gamma)\right]$ through a hidden layer of resonant modes, which act as neural units with activation function $\boldsymbol{\sigma}$ and output weights $\boldsymbol{\beta}$ (Fig. \ref{funo}d).\\
    The activation $\boldsymbol\sigma=\frac{\mb K}{i(\omega-\boldsymbol\Omega)+\boldsymbol\Gamma}$ is a rational function (see Supplementary Note I) and satisfies the conditions for the universal approximation theorem of neural networks~\cite{LESHNO1993861}: the function is non-polynomial and possesses the complex threshold parameters $\boldsymbol\Gamma-i\boldsymbol{\Omega}$. This implies that electromagnetic resonances can be used as universal approximators: Suppl. Note II rigorously demonstrates in the general case that by controlling the position of $M$
    resonant frequencies in $\boldsymbol\Omega$, regardless of the values of $\boldsymbol\Gamma$, $\mb C$ and $\mb K$, it is possible to exactly set the
    output response $\mb H(\omega)$ in amplitude and phase at $2L\le M-1$  frequency points $\omega_l$ ($l=1,...,L$), and in least-square sense at a number of spectral points $2L\ge M$, in any desired scattering channel.\\
    Figure \ref{funo}e-h
    illustrates this point quantitatively. We considered an
    optical network initialized to random values of $\boldsymbol\Omega$, $\boldsymbol\Gamma$, $\mb C$ and $\mb K$, and
    set the real and imaginary part of the output transmission
    $s_{+1}^{(target)}=T_{real}+iT_{imag}$ of one scattering channel to random values at $L=5$ different
    spectral points (Fig. \ref{funo}e-f, red markers). Initialization details are in Suppl. Note III. By using an iterative
    optimization, we trained the
    resonant frequencies $\boldsymbol\Omega$ to approximate the desired response by minimizing the cost
    function $F=\norm{s_{+1}^{(target)}-s_{+1}^{(predicted)}}$, with
    $s_{+1}^{(predicted)}$ being the output generated by the optical network with
    trained $\boldsymbol\Omega$. In the optimization
    procedure we changed only the resonance frequencies $(\boldsymbol\Omega)_{nn}$, without altering $\mb K$,
     $\boldsymbol\Gamma$ or $\mb C$.\\
    Figure \ref{funo}g shows the value of the cost
    function for the best network obtained at increasing number of modes $M$.
    Once the mode number becomes larger than $2L+1$, in agreement with the results of Suppl. Note II, the network represents the
    desired response (Fig. \ref{funo}e-f, solid lines). Figure \ref{funo}h
    visualizes the resonances network with $M=12$, shown as a connected graph with nodes at
     $(\Omega_n,\Gamma_{nn}\equiv\gamma_n)$, and  links between nodes
    $n$ and $m$ representing the modes coupling strength $|\Gamma_{nm}|$. \\
    By using universal approximators, the design of flat-optics  is  reduced to learning a suitable set of resonant frequencies in the hidden layer of Fig. \ref{funo}d. These frequencies are learnt by geometrical deformations: As an example, for a cuboid optical
    dielectric resonator of refractive index $n_{r}$ and dimensions
    $L_x$, $L_y$, $L_z$ terminated by PMC boundary conditions, the resonant
    frequencies $\Omega_{nmp}$ are: \begin{equation} \label{res0}
    \Omega_{nmp}=\frac{c}{n_{r}}\sqrt{\left(\frac{n\cdot\pi}{L_x}\right)^2
	+\left(\frac{m\cdot\pi}{L_y}\right)^2+\left(\frac{p\cdot\pi}{L_z}\right)^2},
    \end{equation} and can be adjusted by deforming the resonator shape via
    $L_x$, $L_y$, and $L_z$. Figure \ref{funo}i reports the number of
    resonances at visible wavelengths between 300~nm and 800~nm, contained in a Silicon (Si)
    cuboid resonator of thickness $L_z=50$~nm with variable dimensions $L_x$ and
    $L_y$, ranging from $50$~nm to $500$~nm. Subwavelength cuboid
    resonators contain hundreds of resonances, allowing the design of universal approximators for a broad range of desired frequency responses. Other shapes (e.g., cylindrical, spherical,...) provide similar results. In this work we focus on cuboid shapes for their fabrication versatility.\\ 
The main challenge in engineering Eq. \eqref{res0} is the nonlinear relationship between the activation function $\sigma=\frac{\mb K}{i(\omega-\boldsymbol\Omega)+\boldsymbol\Gamma}$ and the resonant frequencies $\boldsymbol\Omega$: modifying the
geometry of one resonator via \eqref{res0}, modifies the response of all the others. This requires carrying out the search by simultaneously optimizing all the frequencies of the network. To perform this task, we developed an Autonomous Learning Framework for Rule-based Evolutionary Design
(ALFRED) software, compatible with parallel supercomputer architectures. \\
ALFRED consists of two main parts: an optimizer and a predictor (Fig.
\ref{fdue}a). The global search for the best configuration of resonances is
carried out by a swarm particle optimizer, which is effective in training
high-dimensional neural networks~\cite{10.5555/2876238,Kennedy:2001:SI:370449}. The swarm
performs a collective search based on an ensemble of randomly defined tentative particle solutions (Fig. \ref{fdue}a), with each particle representing a specific geometry of Si boxes resonators.\\
The swarm algorithm
developed is a parallel version of~\cite{Iadevaia6704,5397662}, in which the search parameters (inertia, social,
and memory in Fig. \ref{fdue}a) are autonomously adapted from the
collective interactions in the swarm. The cost function is defined as the norm between the desired response and the predicted one from the particle under consideration. In our implementation, the single particle evolution is
carried out by a CPU-core of a supercomputer architecture, and particles
interactions are carried out in parallel between different cores, thus 
speeding up the global search.\\
The main bottleneck of
the optimizer is the time required to evaluate the cost function: each particle needs to calculate the
output response from the structure by using first principle simulations, which are essential
to keep into account all material effects (e.g., dispersion,...)
that can furnish a precise design.
For this reason, we added to each particle a neural network predictor unit
 (Fig. \ref{fdue}b), which is trained by finite-difference time-domain (FDTD) to predict the outcome of first principle simulations.\\
The predictor is designed with a convolutional neural
network (CNN) based on a ResNet18 architecture~\cite{resnet}, which
operates on an image representing a single cell of the structure, followed by a series of fully connected neural networks (FCN). The CNN extracts multidimensional features from the image and feeds them to FCN layer, which is trained to predict the output
response $\mb s_o(\omega)$ from the features extracted by the CNN layer. We trained different FCN with discrete thicknesses in the range from 50 to 300~nm with a step of 25~nm, and connected them sequentially to the output from a CNN block through a logical
switch (Fig. \ref{fdue}b, S), which chooses the appropriate  block.\\
The training dataset for the predictor is composed by arrays of Si boxes structures (currently, up to five), and is self-generated by ALFRED and autonomously optimized in background. This is
accomplished by mapping the dataset into a 
multidimensional features space, generating additional datasets in the regions
where the predictions are lower than a predetermined threshold ($80\%$).\\
The results of Eqs. \eqref{res0} show that in order to reach the correct solution, the system has to be able to explore the space of all possible deformations in the resonators shape. To comply with this condition, ALFRED does not make any assumption on the system periodicity, and optimize this value autonomously.\\ 
Supplementary Fig. 1 shows a typical prediction example with the relative
dataset. The predictions match FDTD computations for both TE and TM input polarizations, with more than 99\% of predictions above threshold.
During a typical search, ALFRED first uses the swarm equipped with  the CNN+FCN predictor to rapidly converge to an initial structure, and then removes the predictor and launches a final 
swarm optimization with parallel
FDTD simulations to
validate the exact design.\\
Figure \ref{fdue}c-f illustrates an example design obtained by ALFRED for a polarizing beam splitter at the operating wavelength of $900$~nm. The cost function $F$ (Fig. \ref{fdue}c) minimized is defined as follows:
\begin{equation}
    \label{cost0}
    F=\norm{1-s_{+}^{TE}(\omega_0)}+\norm{1-s_{-}^{TM}(\omega_0)}+\norm{\frac{d s_{+}^{TE}(\omega_0)}{d\omega}}+\norm{\frac{ds_{-}^{TM}(\omega_0)}{d\omega}},
\end{equation}
with $s_{+}^{TE}(\omega_0)$ and $s_{-}^{TM}(\omega_0)$ being the transmission and reflection measured on a flat scattering wavefront at the operating frequency $\omega_0=2\pi c/\lambda_0$ for TE and TM polarizations, respectively, and $\norm{.}$ the norm. The cost function $F$ maximizes the transmission for TE and the reflection for TM on a flat wavefront and provides broadband performances by minimizing the first derivative in transmission and reflection. Broader performances can be obtained by minimizing higher order derivatives, which provide more flat frequency responses.\\
Figure \ref{fdue}d show the values of the cost function minimized by ALFRED during progressive iterations. The final structure  (Supplementary Fig. 2), is composed of four Si boxes of $100$~nm thickness, arranged in a complex geometrical pattern. The cost functions reaches values below $10^{-3}$, which result in a device efficiency of $99.99\%$. Figure \ref{fdue}f illustrates this result quantitatively with FDTD simulations for the TE and TM polarizations at $\omega_0$, showing the behavior of the structure with the generation of completely flat wavefronts in both reflection and transmission, without any wavefront aberration.
\subsubsection*{Experiments: polarizing beam splitters, dichroic mirrors and two sub-pixel color displays}
Using ALFRED we design, fabricate, and characterize different structures for different applications. Once model is trained, numerical simulations are performed on a single GPU, for a total simulation time of a few minutes. A strength of this technique is the use of the same manufacturing (see Methods) for all devices.\\ 
We considered a first series of devices for polarization control in both transmission and reflection, on a challenging component that is currently tackled by both direct and inverse designs, allowing a quantitative comparison with our approach. \Cref{Polarizer_characterisation}a summarizes the
current state of the art of flat optics polarizers. Currently, existing designs provide polarization filtering in transmission, with the best reported efficiency being around $65\%$ in the visible range (direct design approach) and around $77\%$ for infrared wavelengths (inverse design approach).\\
We use ALFRED to design ultra-flat polarizing beam splitters centered at different laser line frequencies, with full polarization control in reflection and in transmission. The designs are obtained by the minimizing same fitness function of Eq. \eqref{cost0}, illustrated in Fig. \ref{fdue}c-d, and by centering the wavelength $\lambda_0$ around \SIlist{533; 600; 750}{nm}. In all cases, the results obtained are the same of Fig. \ref{fdue}c-d: the fitness function is reduced to values below $10^{-3}$, with theoretical transmission and reflection efficiencies beyond $99\%$, including material losses, in structures with thickness of $\Delta t=50$~nm ($\lambda_0=533$~nm and $\lambda_0=750$~nm) and $\Delta t=56$~nm ($\lambda_0=600$~nm), represented by different periodic configurations of Si boxes, with periods $\Lambda=290$~nm, $\Lambda=344$~nm, and $\Lambda=482$~nm, for polarizers at $\lambda_0=533$~nm, $\lambda_0=600$~nm and $\lambda_0=750$~nm, respectively. Figure \ref{fdue}e shows the geometry of a representative sample for $\lambda_0=900$~nm. Key to obtain these performances is the simultaneous optimization of several parameters on the flat wavefront from an initial design using universal approximators, motivated by the theoretical understanding developed in the previous section.\\  
We experimentally assess the performances of $2$~mm~$\times$~$2$~mm fabricated devices by the setup of Fig. \ref{Polarizer_characterisation}b. We illuminate each sample with an NKT Photonics SuperK EXTREME supercontinuum laser source (Fig. \ref{Polarizer_characterisation}b, SL), followed by a broadband linear polarizer
mounted on a computer controlled rotating stage (Fig. \ref{Polarizer_characterisation}b, MLP), and then measure transmission and reflection for each wavelength and for each angular orientation $\Delta\theta$ of the reference polarizer with two calibrated photodetectors (Fig. \ref{Polarizer_characterisation}b, PD).\\
\Cref{Polarizer_characterisation}c summarizes the polarization efficiency results for each sample and for each wavelength, while Fig. \ref{Polarizer_characterisation}d shows scanning electron microscope images (SEM) of a representative sample. The polarization efficiency, defined
as $\eta=\qty(\frac{T_{max}-T_{min}}{T_{max}+T_{min}})$~\cite{Bass2010}, is evaluated as $\eta=97\%$,
$\eta=99\%$, $\eta=96\%$ for the \SIlist{533; 600; 750}{nm} polarizing beam splitters in the visible, respectively, and $\eta=88\%$ in the nearIR for the  $\lambda_0=900$ polarizer. 
Figure \ref{Polarizer_characterisation}e reports the result of  transmission measurements for the \SI{600}{nm} polarizing beam splitter, showing a transmission beyond 95\% over a large portion of the visible spectrum when aligned with the analyzer
and a reflection of over 95\% of the light in a 75~nm range centered at the wavelength of 595~nm when perpendicular to the analyzer. Figure \ref{Polarizer_characterisation}f displays theoretical (solid line) and experimental (dashed line) retrieved spectra of the polarizer at $600$~nm, illustrating good agreement between designed and measured performances. Supplementary Fig. 2 shows the experimentally measured polarization behavior at the 600~nm wavelength, showing the expected Malus's law sinusoidal curve~\cite{malus}.
The near unity performance of the polarizers designed at visible frequencies relies on the ultra-thin thickness of the Si structure, which operates as a transparent layer with negligible losses.\\
Figure \ref{Dichroic_mirror} reports the design and characterization of a dichroic mirror, which acts as a wavelength demultiplexer by
transmitting light at one frequency $\omega_0$ and reflecting light at a different frequency $\omega_1\neq \omega_0$. We choose this example as it represents a device that is not yet developed and that has a broadband response defined over hundreds of nm, representing an interesting broadband benchmark for ALFRED.\\ 
The cost function for this device is defined as follows:
\begin{equation}
    \label{dicr0}
 F=\norm{1-s_{+}^{T}(\omega_0)}+\norm{1-s_{-}^{R}(\omega_1)}+\norm{\frac{d s_{+}^{T}(\omega_0)}{d\omega}}+\norm{\frac{ds_{-}^{R}(\omega_1)}{d\omega}},
\end{equation}
with $s_{+}^T(\omega)$ and $s_{-}^R(\omega)$ being the transmission and reflection of the structure measured on a flat scattering wavefront at the frequency $\omega$. Figure \ref{Dichroic_mirror}a reports the results of ALFRED for a swarm of 40 particles, showing a reduction of the cost function $F$ to values below $10^{-1}$ after $60$ iterations. The design structure found by ALFRED is composed of an aperiodic pattern of lines with thickness $\Delta t=209$~nm, period $\Lambda=542$~nm, and widths $250$~nm and $40$~nm (Fig. \ref{Dichroic_mirror}b). This results also shows the ability of ALFRED to explore a large manifold of solutions, converging automatically to diverse types of 2D/3D structures that can solve the problem with high efficiency.\\ 
Figure \ref{Dichroic_mirror}c (dashed lines) shows the FDTD calculated spectral response of the structure, showing the required behavior around the target points (Fig. \ref{Dichroic_mirror}b, circle markers) on a bandwidth larger than $250$~nm, with transmission efficiencies above $95\%$ and a rejection below $1\%$ in reflection.\\
The experimental response of a fabricated sample (Fig. \ref{Dichroic_mirror}d) is reported in Fig. \ref{Dichroic_mirror}c (solid lines). Experimental results match quite well with the theory on all the frequency range considered, reporting an experimental efficiency of $90\%$ in transmission and a reflection in the range of few percents.\\
At variance with a traditional dichroic, which requires a  macroscopic thickness due to the $45^\circ$ wavelength mixing geometry (Fig. \ref{Dichroic_mirror}e), the flat-optics sample of Fig. \ref{Dichroic_mirror}d performs the same functionality at normal incidence and in a structure of only $\approx 200$~nm thickness, allowing flat system integration.\\
Figure \ref{pixel2-figure} summarizes the design results for a metasurface color display based on a two sub-pixel technology, comparing it with the current state of the art. This last example targets an inverse designed functionality defined over all the visible bandwidth. In current LCD displays, colors result from an unpolarized broadband back light (Fig. \ref{pixel2-figure}a, BL), which is modulated in
intensity via a liquid crystal cell (Fig. \ref{pixel2-figure}a, LC) equipped with two orthogonal linear
polarizers (Fig. \ref{pixel2-figure}a, LP), and then filtered into primary red, green, and blue components (Fig. \ref{pixel2-figure}a, RGB). In displays based on organic LED's, the colors are directly produced by organic monochromatic emitters at different frequencies, which are independently controlled (Fig. \ref{pixel2-figure}b). In both displays, composite colors are obtained by controlling in intensity the three primary components.\\
We design an integrated architecture (Fig. \ref{pixel2-figure}c), which allows polarization-intensity gamut control with only two sub-pixels. The unit cell uses two independently controlled back light sources, followed by a polarization stage composed, as in LCD displays, of a linear polarizer and a liquid crystal cell (Fig. \ref{pixel2-figure}c, LP+LC), in which the output light polarization state is rotated. The final colors are produced by employing two flat-optics polarization filters $A$ and $B$, which output different chromaticities when the input light polarization changes orientation.\\
By generating RGB composite colors from two sub-pixels, this architecture allows reducing the energy consumption of a traditional screen by $33\%$, and for the same screen size, it allows to increase the resolution by $33\%$. To the best of our knowledge, no flat optics structure has been designed to address this functionality yet.\\
To design the required structures, we maximized the following cost function $F$:
\begin{equation}
    \label{displ0}
    F=\mathrm{Area_{gamut}}\left[A_{pol,int},B_{pol,int}\right],
\end{equation}
representing the gamut area of all the possible chromaticities obtained by combining different polarizations and intensities at the input of samples $A$ and $B$. To perform this calculation, we convert the transmitted electromagnetic spectra for samples $A$ and $B$ illuminated by TE/TM polarizations into $xy$ chromaticity coordinates of a standard CIE 1931 chromaticity diagram~\cite{10.5555/3051718}. The gamut of all possible chromaticities is the area of the rhomboid whose four vertices are the $xy$ coordinates $TE_{A},TM_{A},TE_{B},TM_{B}$ of the spectra obtained from the two samples, $A$ and $B$, under TE and TM polarization respectively. This results from the condition that the combination in intensity of two chromaticity coordinates generates all the possible points on the line connecting the two initial coordinates~\cite{doi:10.1002/col.5080080421}.\\
Figure \ref{pixel2-figure}d-e illustrates the results of two set of optimal structures found by ALFRED. The first design is composed by $A$ and $B$ samples made by periodic cells containing one box each (Fig. \ref{pixel2-figure}d, gray/brown unit cells), with thicknesses $\Delta t=170$~nm and $\Delta t=218$~nm, respectively. When the polarization changes from TE to TM, sample $A$ represents blue-red colors (Fig. \ref{pixel2-figure}d, $TE_A-TM_A$ line), while sample $B$ generates green to blue (Fig. \ref{pixel2-figure}d, $TE_B-TM_B$ line). The second sample configuration (Fig. \ref{pixel2-figure}e) is composed of unit cells with complex two and three configurations of boxes possessing thickness $\Delta t=225$~nm. This set of samples provides an improved performance, exhibiting a larger color gamut that even exceeds the standard RGB spectrum (Fig. \ref{pixel2-figure}d, dashed yellow line).\\
We demonstrate the proof of concept of this technology by manufacturing the optimized structure in Fig. \ref{pixel2-figure}d, whose chromaticity gamut lies mostly within the RGB spectrum and can be visualized with a standard RGB camera. We characterized the performance of the samples by using the setup of Fig. \ref{Polarizer_characterisation}b, in which the MLP stage mimics the role of the liquid crystal cell, using the illumination of a white LED source in place of the supercontinuum laser.\\ 
Figure \ref{pixel-figure}a compares theoretical (black rhomboid) and experimental (blue rhomboid) gamut curves, obtained from manufactured samples $A$ and $B$ (Fig. \ref{pixel-figure}b). Different colors are reported for each sample and for the varying polarization angle $\Delta\phi$ of the electric field $\mb E$, varied between $0^\circ$ and $90^\circ$, whose extrema represent TM and TE polarization, respectively. Figure \ref{pixel-figure}c reports the corresponding measured spectra (solid lines), compared to the theoretical predictions from ALFRED (solid lines). The experimental results match the theory with good agreement, both in the spectral response (Fig. \ref{pixel2-figure}c) and in the experimentally retrieved colors (Fig. \ref{pixel-figure}b), resulting in a good representation of the theoretical gamut of colors inside the RGB triangle of Fig. \ref{pixel-figure}a. The very small differences in the measured spectra for wavelengths larger than $600$~nm in Fig. \ref{pixel-figure}c generates the small deviations observed in Fig. \ref{pixel-figure}a in the blue and black curves. Figure \ref{pixel-figure}c shows that both samples $A$ and $B$ have almost ideal, near unity, maximum transmission values. 

\section*{Conclusions}
We discussed the inverse design of flat optics by exploiting a hidden network of universal approximators engineered in suitably deformed dielectric nanoresonators, demonstrating various components for vectorial light control with near unity performances. This technology controls simultaneously the device transmission, reflection and the desired output wavefront shape, with no distortion, and does not employ predefined assumptions on the system periodicity, converging automatically to globally optimized solutions characterized by high efficiency values over large design bandwidths. These results open the doors for the realization of integrated surfaces for general light processing with high efficiency at visible frequencies, and for integrated optics in ultra-flat surfaces with comparable efficiency to their traditional bulk counterparts.

\section*{Methods}

\subsection{Sample Nanofabrication} 
We used as a base wafer a square piece of borosilicate glass, \SI{18}{mm} wide and
approximately \SI{200}{\micro\meter} thick (12-540-A, from Fisher
scientific), which is cleaned by acetone and isopropyl alcohol. We then grow 
a uniform layer of amorphous silicon via plasma enhanced chemical vapour
deposition (PECVD). We control the thickness of the Si layer by using ellipsometry (UVISEL Plus, from HORIBA). We then spin coat
a positive electron beam resist ZEP 520A (from ZEON corporation) on the sample at
\SI{4000}{RPM} for \SI{60}{s}, after which we bake it on a hotplate at
\SI{180}{\celsius} for \SI{3}{\minute}. After this step, we spin coat a
conductive polymer AR-PC 5090.02 (ALLRESIST)  onto the sample at \SI{4000}{RPM}
for \SI{60}{s} and bake the device again on a hotplate for \SI{1}{\minute} at
\SI{100}{\celsius}. We write the optical resonators pattern by using a JEOL JBX-6300FS electron beam
lithography system at a \SI{100}{\kilo\volt} accelerating voltage. After writing, we remove the polymer by submerging the sample in deionized water for \SI{60}{s}, and develop the resist with n-Amyl acetate (ZED-N50 from the ZEON
corporation) for \SI{90}{s} and by submersion in isopropyl alcohol for
\SI{90}{s}. We then use electron beam evaporation to deposit a \SI{22}{nm} layer
of chromium on the sample. We perform liftoff by submerging the sample
in N-methylpyrrolidone (ALLRESIST) at \SI{70}{\celsius} for one hour and
sonicate the solution for one minute afterwards to create a protective mask in the image of the resonator pattern that we intend to fabricate. We
then use reactive ion etching with $\text{SF}_6$ to remove the unprotected silicon and
expose the underlying glass. We remove the chromium mask by
submersion in a perchloric acid and ceric ammounium nitrate solution (TechniEtch
Cr01 from MicroChemicals) for \SI{30}{\second}.



\begin{addendum} \item This research received funding from KAUST (Award
    OSR-2016-CRG5-2995). Parallel simulations are performed on KAUST's Shaheen supercomputer.  \item[Competing Interests] A provisional patent
	application  (US Patent Application No. 62/844,416) has been filed on
	the basis of the results of this work with all authors of the manuscript listed as inventors.
 
 \item[Correspondence] Correspondence and requests for materials should be
addressed to Andrea Fratalocchi (andrea.fratalocchi@kaust.edu.sa)

\end{addendum}
\printbibliography

@article{Guo2017b,
  title = {High-{{Order Dielectric Metasurfaces}} for {{High}}-{{Efficiency Polarization Beam Splitters}} and {{Optical Vortex Generators}}},
  author = {Guo, Zhongyi and Zhu, Lie and Guo, Kai and Shen, Fei and Yin, Zhiping},
  date = {2017-08-29},
  journaltitle = {Nanoscale Research Letters},
  shortjournal = {Nanoscale Research Letters},
  volume = {12},
  pages = {512},
  issn = {1556-276X},
  doi = {10.1186/s11671-017-2279-2},
  number = {1}
}

@article{Guo2007,
  title = {Nanoimprint {{Lithography}}: {{Methods}} and {{Material Requirements}}},
  shorttitle = {Nanoimprint {{Lithography}}},
  author = {Guo, L. J.},
  date = {2007-02-19},
  journaltitle = {Advanced Materials},
  shortjournal = {Adv. Mater.},
  volume = {19},
  pages = {495--513},
  issn = {09359648, 15214095},
  doi = {10.1002/adma.200600882},
  langid = {english},
  number = {4}
}

@collection{Gendreau2019,
  title = {Handbook of {{Metaheuristics}}},
  editor = {Gendreau, Michel and Potvin, Jean-Yves},
  date = {2019},
  volume = {272},
  publisher = {{Springer International Publishing}},
  location = {{Cham}},
  doi = {10.1007/978-3-319-91086-4},
  isbn = {978-3-319-91085-7},
  langid = {english},
  series = {International {{Series}} in {{Operations Research}} \& {{Management Science}}}
}

@article{Johnson2014,
  title = {A Brief Review of Atomic Layer Deposition: From Fundamentals to Applications},
  shorttitle = {A Brief Review of Atomic Layer Deposition},
  author = {Johnson, Richard W. and Hultqvist, Adam and Bent, Stacey F.},
  date = {2014},
  journaltitle = {Materials Today},
  shortjournal = {Materials Today},
  volume = {17},
  pages = {236--246},
  publisher = {{Elsevier}},
  issn = {13697021},
  doi = {10.1016/j.mattod.2014.04.026},
  number = {5}
}

@article{Malkiel2018,
  title = {Plasmonic Nanostructure Design and Characterization via {{Deep Learning}}},
  author = {Malkiel, Itzik and Mrejen, Michael and Nagler, Achiya and Arieli, Uri and Wolf, Lior and Suchowski, Haim},
  date = {2018-09-05},
  journaltitle = {Light: Science \& Applications},
  shortjournal = {Light Sci Appl},
  volume = {7},
  pages = {1--8},
  publisher = {{Nature Publishing Group}},
  issn = {2047-7538},
  doi = {10.1038/s41377-018-0060-7},
  issue = {1},
  langid = {english},
  number = {1}
}

@article{Andkjaer2014,
  title = {Inverse Design of Nanostructured Surfaces for Color Effects},
  author = {Andkj\ae{}r, Jacob and Johansen, Villads Egede and Friis, Kasper Storgaard and Sigmund, Ole},
  date = {2014-01-01},
  journaltitle = {JOSA B},
  shortjournal = {J. Opt. Soc. Am. B, JOSAB},
  volume = {31},
  pages = {164--174},
  publisher = {{Optical Society of America}},
  issn = {1520-8540},
  doi = {10.1364/JOSAB.31.000164},
  langid = {english},
  number = {1}
}

@inproceedings{Frandsen2016,
  title = {Inverse Design Engineering of All-Silicon Polarization Beam Splitters},
  booktitle = {Photonic and {{Phononic Properties}} of {{Engineered Nanostructures VI}}},
  author = {Frandsen, Lars H. and Sigmund, Ole},
  date = {2016-03-14},
  volume = {9756},
  pages = {97560Y},
  publisher = {{International Society for Optics and Photonics}},
  doi = {10.1117/12.2210848},
  eventtitle = {Photonic and {{Phononic Properties}} of {{Engineered Nanostructures VI}}}
}

@article{Lin2020,
  title = {Inverse Design of Plasmonic Metasurfaces by Convolutional Neural Network},
  author = {Lin, Ronghui and Zhai, Yanfen and Xiong, Chenxin and Li, Xiaohang},
  date = {2020-03-15},
  journaltitle = {Optics Letters},
  shortjournal = {Opt. Lett., OL},
  volume = {45},
  pages = {1362--1365},
  publisher = {{Optical Society of America}},
  issn = {1539-4794},
  doi = {10.1364/OL.387404},
  langid = {english},
  number = {6}
}

@article{Liu2018b,
  title = {Training {{Deep Neural Networks}} for the {{Inverse Design}} of {{Nanophotonic Structures}}},
  author = {Liu, Dianjing and Tan, Yixuan and Khoram, Erfan and Yu, Zongfu},
  date = {2018-04-18},
  journaltitle = {ACS Photonics},
  shortjournal = {ACS Photonics},
  volume = {5},
  pages = {1365--1369},
  publisher = {{American Chemical Society}},
  doi = {10.1021/acsphotonics.7b01377},
  number = {4}
}

@article{Nadell2019,
  title = {Deep Learning for Accelerated All-Dielectric Metasurface Design},
  author = {Nadell, Christian C. and Huang, Bohao and Malof, Jordan M. and Padilla, Willie J.},
  date = {2019-09-30},
  journaltitle = {Optics Express},
  shortjournal = {Opt. Express, OE},
  volume = {27},
  pages = {27523--27535},
  publisher = {{Optical Society of America}},
  issn = {1094-4087},
  doi = {10.1364/OE.27.027523},
  langid = {english},
  number = {20}
}

@article{Phan2019,
  title = {High-Efficiency, Large-Area, Topology-Optimized Metasurfaces},
  author = {Phan, Thaibao and Sell, David and Wang, Evan W. and Doshay, Sage and Edee, Kofi and Yang, Jianji and Fan, Jonathan A.},
  date = {2019-05-29},
  journaltitle = {Light: Science \& Applications},
  shortjournal = {Light Sci Appl},
  volume = {8},
  pages = {1--9},
  publisher = {{Nature Publishing Group}},
  issn = {2047-7538},
  doi = {10.1038/s41377-019-0159-5},
  issue = {1},
  langid = {english},
  number = {1}
}

@article{SeanMolesky2018,
  title = {Inverse Design in Nanophotonics},
  author = {{Sean  Molesky} and {Zin   Lin} and {Alex  Piggott} and {er Y.} and {Weiliang   Jin} and {Jelena   Vuckovic} and {Alej  Rodriguez} and {ro W.}},
  date = {2018},
  journaltitle = {1801.06715},
  shortjournal = {Nat Photonics},
  volume = {12},
  pages = {659--670},
  issn = {1749-4885},
  doi = {10.1038/s41566-018-0246-9},
  number = {11}
}

@article{Sell2017a,
  title = {Large-{{Angle}}, {{Multifunctional Metagratings Based}} on {{Freeform Multimode Geometries}}},
  author = {Sell, David and Yang, Jianji and Doshay, Sage and Yang, Rui and Fan, Jonathan A.},
  date = {2017-06-14},
  journaltitle = {Nano Letters},
  shortjournal = {Nano Lett.},
  volume = {17},
  pages = {3752--3757},
  publisher = {{American Chemical Society}},
  issn = {1530-6984},
  doi = {10.1021/acs.nanolett.7b01082},
  number = {6}
}

@article{Sell2017b,
  title = {Periodic {{Dielectric Metasurfaces}} with {{High}}-{{Efficiency}}, {{Multiwavelength Functionalities}}},
  author = {Sell, David and Yang, Jianji and Doshay, Sage and Fan, Jonathan A.},
  date = {2017},
  journaltitle = {Advanced Optical Materials},
  volume = {5},
  pages = {1700645},
  issn = {2195-1071},
  doi = {10.1002/adom.201700645},
  langid = {english},
  number = {23}
}

@article{Shen2014,
  title = {Ultra-High-Efficiency Metamaterial Polarizer},
  author = {Shen, Bing and Wang, Peng and Polson, Randy and Menon, Rajesh},
  date = {2014-11-20},
  journaltitle = {Optica},
  shortjournal = {Optica, OPTICA},
  volume = {1},
  pages = {356--360},
  publisher = {{Optical Society of America}},
  issn = {2334-2536},
  doi = {10.1364/OPTICA.1.000356},
  langid = {english},
  number = {5}
}

@article{Shen2015,
  title = {An Integrated-Nanophotonics Polarization Beamsplitter with 2.4 \texttimes{} 2.4 {$M$}m 2 Footprint},
  author = {Shen, Bing and Wang, Peng and Polson, Randy and Menon, Rajesh},
  date = {2015-06},
  journaltitle = {Nature Photonics},
  shortjournal = {Nature Photon},
  volume = {9},
  pages = {378--382},
  publisher = {{Nature Publishing Group}},
  issn = {1749-4893},
  doi = {10.1038/nphoton.2015.80},
  issue = {6},
  langid = {english},
  number = {6}
}

@article{Singh2019,
  title = {Inverse {{Design}} of {{Photonic Metasurface Gratings}} for {{Beam Collimation}} in {{Opto}}-Fluidic {{Sensing}}},
  author = {Singh, Robin and Nie, Yuqi and Agarwal, Anuradha Murthy and Anthony, Brian W.},
  date = {2019-10-30},
  url = {http://arxiv.org/abs/1911.08957},
  urldate = {2020-04-09},
  archivePrefix = {arXiv},
  eprint = {1911.08957},
  eprinttype = {arxiv},
  langid = {english},
  primaryClass = {physics}
}

@article{Pelletier2006,
  title = {Aluminum Nanowire Polarizing Grids: {{Fabrication}} and Analysis},
  shorttitle = {Aluminum Nanowire Polarizing Grids},
  author = {Pelletier, Vincent and Asakawa, Koji and Wu, Mingshaw and Adamson, Douglas H. and Register, Richard A. and Chaikin, Paul M.},
  date = {2006-05-22},
  journaltitle = {Applied Physics Letters},
  shortjournal = {Appl. Phys. Lett.},
  volume = {88},
  pages = {211114},
  issn = {0003-6951},
  doi = {10.1063/1.2206100},
  number = {21}
}

@article{Wang2007d,
  title = {Polarizing Beam Splitter of a Deep-Etched Fused-Silica Grating},
  author = {Wang, Bo and Zhou, Changhe and Wang, Shunquan and Feng, Jijun},
  date = {2007-05-15},
  journaltitle = {Optics Letters},
  shortjournal = {Opt. Lett.},
  volume = {32},
  pages = {1299},
  issn = {0146-9592, 1539-4794},
  doi = {10.1364/OL.32.001299},
  langid = {english},
  number = {10}
}

@article{Hong2007b,
  title = {Silicon Nanowire Grid Polarizer for Very Deep Ultraviolet Fabricated from a Shear-Aligned Diblock Copolymer Template},
  author = {Hong, Young-Rae and Asakawa, Koji and Adamson, Douglas H. and Chaikin, Paul M. and Register, Richard A.},
  date = {2007-11-01},
  journaltitle = {Optics Letters},
  shortjournal = {Opt. Lett.},
  volume = {32},
  pages = {3125},
  issn = {0146-9592, 1539-4794},
  doi = {10.1364/OL.32.003125},
  langid = {english},
  number = {21}
}

@article{2002.08121,
  title = {Generalized {{Maxwell}} Projections for Multi-Mode Network {{Photonics}}},
  author = {Makarenko, M. and Burguete-Lopez, A. and Getman, F. and Fratalocchi, A.},
  year = {2020},
  url = {http://arxiv.org/abs/2002.08121},
  urldate = {2020-02-21},
  archivePrefix = {arXiv},
  eprint = {2002.08121},
  eprinttype = {arxiv},
  primaryClass = {physics},
  note={[To be published]}
}

@article{10.1093/nsr/nwy017,
  title = {All-Dielectric Meta-Optics and Non-Linear Nanophotonics},
  author = {Kivshar, Yuri},
  date = {2018-01},
  journaltitle = {National Science Review},
  volume = {5},
  pages = {144--158},
  issn = {2095-5138},
  doi = {10.1093/nsr/nwy017},
  eprint = {https://academic.oup.com/nsr/article-pdf/5/2/144/31568035/nwy017.pdf},
  number = {2}
}

@book{10.5555/17318,
author = {Braess, Dietrich},
title = {Nonlinear Approximation Theory},
year = {1986},
isbn = {0387136258},
publisher = {Springer-Verlag},
address = {Berlin, Heidelberg}
}

@book{malus,
author = {Edward Coffett},
title = {Field Guide to Polarization},
year = {2005},
publisher = {SPIE},
isbn = {9780819458681},
pages={148}
}

@article{Totero_Gongora_2017,
	doi = {10.1088/1361-6528/aa593d},
	url = {https://doi.org/10.1088%2F1361-6528%2Faa593d},
	year = 2017,
	month = {2},
	publisher = {{IOP} Publishing},
	volume = {28},
	number = {10},
	pages = {104001},
	author = {Juan Sebastian Totero Gongora and Gael Favraud and Andrea Fratalocchi},
	title = {Fundamental and high-order anapoles in all-dielectric metamaterials via Fano{\textendash}Feshbach modes competition},
	journal = {Nanotechnology}
}

@article{Kirill2019,
  title = {Nonradiating Photonics with Resonant Dielectric Nanostructures},
  author = {Koshelev, Kirill and Favraud, Gael and Bogdanov, Andrey and Kivshar, Yuri and Fratalocchi, Andrea},
  date = {2019-05-27},
  journaltitle = {Nanophotonics},
  volume = {8},
  pages = {725--745},
  issn = {2192-8614},
  doi = {10.1515/nanoph-2019-0024},
  number = {5}
}

@book{10.5555/2876238,
  title = {Multidimensional Particle Swarm Optimization for Machine Learning and Pattern Recognition},
  author = {Kiranyaz, Serkan and Ince, Turker and Gabbouj, Moncef},
  date = {2015},
  edition = {1st},
  publisher = {{Springer Publishing Company, Incorporated}},
  isbn = {3-642-43762-1}
}

@book{10.5555/3051718,
  title = {Encyclopedia of Color Science and Technology},
  author = {Luo, Ronnier},
  date = {2016},
  edition = {1st},
  publisher = {{Springer Publishing Company, Incorporated}},
  isbn = {1-4419-8070-9}
}

@article{1912.07044,
  ids = {Marcucci2019},
  title = {Theory of Neuromorphic Computing by Waves: Machine Learning by Rogue Waves, Dispersive Shocks, and Solitons},
  shorttitle = {Theory of Neuromorphic Computing by Waves},
  author = {Marcucci, Giulia and Pierangeli, Davide and Conti, Claudio},
  date = {2019-12-15},
  url = {http://arxiv.org/abs/1912.07044},
  urldate = {2020-02-11},
  archivePrefix = {arXiv},
  eprint = {arXiv:1912.07044},
  eprinttype = {arxiv}
}

@inproceedings{5397662,
  title = {Parameter Estimation in Dynamic Biochemical Systems Based on Adaptive {{Particle Swarm Optimization}}},
  booktitle = {2009 7th {{International Conference}} on {{Information}}, {{Communications}} and {{Signal Processing}} ({{ICICS}})},
  author = {Liu, Mingshou and Shin, Dongil and Kang, Hwan Il},
  date = {2009-12},
  pages = {1--5},
  publisher = {{IEEE}},
  location = {{Macau, China}},
  doi = {10.1109/ICICS.2009.5397662},
  eventtitle = {Signal {{Processing}} ({{ICICS}})},
  isbn = {978-1-4244-4656-8}
}

@article{Aoni2019,
  title = {High-Efficiency Visible Light Manipulation Using Dielectric Metasurfaces},
  author = {Aoni, Rifat Ahmmed and Rahmani, Mohsen and Xu, Lei and Zangeneh Kamali, Khosro and Komar, Andrei and Yan, Jingshi and Neshev, Dragomir and Miroshnichenko, Andrey E},
  date = {2019-12},
  journaltitle = {Scientific Reports},
  volume = {9},
  pages = {6510},
  issn = {2045-2322},
  doi = {10.1038/s41598-019-42444-y},
  number = {1}
}

@article{Arbabi2015,
  title = {Dielectric Metasurfaces for Complete Control of Phase and Polarization with Subwavelength Spatial Resolution and High Transmission},
  author = {Arbabi, Amir and Horie, Yu and Bagheri, Mahmood and Faraon, Andrei},
  date = {2015-11},
  journaltitle = {Nature Nanotechnology},
  volume = {10},
  pages = {937--943},
  publisher = {{Nature Publishing Group}},
  issn = {17483395},
  doi = {10.1038/nnano.2015.186},
  archivePrefix = {arXiv},
  arxivid = {1411.1494},
  eprint = {1411.1494},
  eprinttype = {arxiv},
  number = {11}
}

@article{BalthasarMueller2017,
  title = {Metasurface {{Polarization Optics}}: {{Independent Phase Control}} of {{Arbitrary Orthogonal States}} of {{Polarization}}},
  shorttitle = {Metasurface {{Polarization Optics}}},
  author = {Balthasar Mueller, J. P. and Rubin, Noah A. and Devlin, Robert C. and Groever, Benedikt and Capasso, Federico},
  date = {2017-03-14},
  journaltitle = {Physical Review Letters},
  shortjournal = {Phys. Rev. Lett.},
  volume = {118},
  pages = {113901},
  doi = {10.1103/PhysRevLett.118.113901},
  number = {11}
}

@book{Bass2010,
  title = {Handbook of Optics. {{Volume I}}, {{Geometrical}} and Physical Optics, Polarized Light, Components and Intruments},
  author = {Bass, Michael and Mahajan, Virendra N.},
  date = {2010},
  edition = {3rd},
  publisher = {{McGraw-Hill}},
  location = {{New York, USA :}},
  isbn = {978-0-07-162925-6}
}

@article{Callewaert2018,
  title = {Inverse-Designed Broadband All-Dielectric Electromagnetic Metadevices},
  author = {Callewaert, F. and Velev, V. and Kumar, P. and Sahakian, A. V. and Aydin, K.},
  date = {2018-12},
  journaltitle = {Scientific Reports},
  volume = {8},
  pages = {1358},
  publisher = {{Nature Publishing Group}},
  issn = {2045-2322},
  doi = {10.1038/s41598-018-19796-y},
  number = {1}
}

@article{Chang2018,
  title = {Optical Metasurfaces: {{Progress}} and Applications},
  author = {Chang, Shengyuan and Guo, Xuexue and Ni, Xingjie},
  date = {2018-07},
  journaltitle = {Annual Review of Materials Research},
  volume = {48},
  pages = {279--302},
  publisher = {{Annual Reviews}},
  issn = {1531-7331},
  doi = {10.1146/annurev-matsci-070616-124220},
  number = {1}
}

@article{Chen2018,
  title = {A Broadband Achromatic Metalens for Focusing and Imaging in the Visible},
  author = {Chen, Wei Ting and Zhu, Alexander Y and Sanjeev, Vyshakh and Khorasaninejad, Mohammadreza and Shi, Zhujun and Lee, Eric and Capasso, Federico},
  date = {2018-01},
  journaltitle = {Nature Nanotechnology},
  volume = {13},
  pages = {220--226},
  publisher = {{Nature Publishing Group}},
  issn = {1748-3387},
  doi = {10.1038/s41565-017-0034-6},
  eprint = {29292382},
  eprinttype = {pmid},
  number = {3}
}

@article{Cheng2016,
  title = {All-Dielectric Ultrathin Conformal Metasurfaces: Lensing and Cloaking Applications at 532 Nm Wavelength},
  author = {Cheng, Jierong and Jafar-Zanjani, Samad and Mosallaei, Hossein},
  date = {2016-12},
  journaltitle = {Scientific Reports},
  volume = {6},
  pages = {38440},
  publisher = {{Nature Publishing Group}},
  issn = {2045-2322},
  doi = {10.1038/srep38440},
  number = {1}
}

@article{Chong2016,
  title = {Efficient Polarization-Insensitive Complex Wavefront Control Using Huygens' Metasurfaces Based on Dielectric Resonant Meta-Atoms},
  author = {Chong, Katie E. and Wang, Lei and Staude, Isabelle and James, Anthony R. and Dominguez, Jason and Liu, Sheng and Subramania, Ganapathi S. and Decker, Manuel and Neshev, Dragomir N. and Brener, Igal and Kivshar, Yuri S.},
  date = {2016-04},
  journaltitle = {ACS Photonics},
  volume = {3},
  pages = {514--519},
  publisher = {{American Chemical Society}},
  issn = {2330-4022},
  doi = {10.1021/acsphotonics.5b00678},
  archivePrefix = {arXiv},
  arxivid = {1602.00755},
  eprint = {1602.00755},
  eprinttype = {arxiv},
  number = {4}
}

@article{Colburn2014,
  title = {Broadband Transparent and {{CMOS}}-Compatible Flat Optics with Silicon Nitride Metasurfaces [{{Invited}}]},
  author = {Colburn, Shane and Zhan, Alan and Bayati, Elyas and Whitehead, James and Ryou, Albert and Huang, Luocheng and Majumdar, Arka},
  date = {2018-08},
  journaltitle = {Optical Materials Express},
  volume = {8},
  pages = {2330},
  issn = {2159-3930},
  doi = {10.1364/OME.8.002330},
  isbn = {2389923909},
  number = {8}
}

@article{Decker2016,
  title = {Resonant Dielectric Nanostructures: A Low-Loss Platform for Functional Nanophotonics},
  author = {Decker, Manuel and Staude, Isabelle},
  date = {2016},
  journaltitle = {Journal of Optics},
  volume = {18},
  pages = {103001},
  publisher = {{IOP Publishing}},
  issn = {2040-8978},
  doi = {10.1088/2040-8978/18/10/103001},
  number = {10}
}

@article{Devlin2016,
  title = {Broadband High-Efficiency Dielectric Metasurfaces for the Visible Spectrum},
  author = {Devlin, Robert C and Khorasaninejad, Mohammadreza and Chen, Wei Ting and Oh, Jaewon and Capasso, Federico},
  date = {2016-09},
  journaltitle = {Proceedings of the National Academy of Sciences},
  volume = {113},
  pages = {10473--10478},
  publisher = {{National Academy of Sciences}},
  issn = {0027-8424},
  doi = {10.1073/pnas.1611740113},
  eprint = {27601634},
  eprinttype = {pmid},
  number = {38}
}

@article{doi:10.1002/adom.201900849,
  title = {Free-Electron Transparent Metasurfaces with Controllable Losses for Broadband Light Manipulation with Nanometer Resolution},
  author = {Bonifazi, Marcella and Mazzone, Valerio and Li, Ning and Tian, Yi and Fratalocchi, Andrea},
  date = {2020},
  journaltitle = {Advanced Optical Materials},
  volume = {8},
  pages = {1900849},
  doi = {10.1002/adom.201900849},
  eprint = {https://onlinelibrary.wiley.com/doi/pdf/10.1002/adom.201900849},
  number = {1}
}

@article{doi:10.1002/col.5080080421,
  title = {Color {{Science}}: {{Concepts}} and {{Methods}}, {{Quantitative Data}} and {{Formulae}}, 2nd Ed., by {{Gunter Wyszecki}} and {{W}}. {{S}}. {{Stiles}}, {{John Wiley}} and {{Sons}}, {{New York}}, 1982, 950 Pp. {{Price}}: \$75.00},
  shorttitle = {Color {{Science}}},
  author = {Billmeyer, Fred W.},
  date = {0024/1983},
  journaltitle = {Color Research \& Application},
  shortjournal = {Color Res. Appl.},
  volume = {8},
  pages = {262--263},
  issn = {03612317, 15206378},
  doi = {10.1002/col.5080080421},
  eprint = {https://onlinelibrary.wiley.com/doi/pdf/10.1002/col.5080080421},
  langid = {english},
  number = {4}
}

@article{frataxai0,
  title = {Ultrafast Integrated Artificial Intelligent Optical Chip. U.S. Pat. Appl. 62/963,747},
  author = {Fratalocchi, A. and Makarenko, M. and Li, X. and Lin, R. and Fariborzi, H. and Alshehri, A. Ali and Yang, K.},
  date = {2020-01},
  publisher = {{Google Patents}},
  url = {http://www.google.it/patents/US4741207}
}

@inproceedings{frataxai2020,
  title = {Artificial Intelligence Inverse Design of Ultra-Flat Meta-Optics with Experimental Efficiencies Exceeding 99\% in the Visible},
  booktitle = {Society of Photo-Optical Instrumentation Engineers ({{SPIE}}) Photonics West Conference Series},
  author = {Burguete-Lopez, Arturo and Makarenko, Maksim and Getman, Fedor and {Andrea Fratalocchi}},
  date = {2020},
  volume = {11299-33}
}

@article{Galinski2017,
  title = {Scalable, Ultra-Resistant Structural Colors Based on Network Metamaterials},
  author = {Galinski, Henning and Favraud, Gael and Dong, Hao and Gongora, Juan S. Totero and Favaro, Grégory and Döbeli, Max and Spolenak, Ralph and Fratalocchi, Andrea and Capasso, Federico},
  date = {2017},
  journaltitle = {Light: Science \& Applications},
  volume = {6},
  pages = {e16233-e16233},
  issn = {2047-7538},
  doi = {10.1038/lsa.2016.233},
  number = {5}
}

@article{Glybovski2016,
  title = {Metasurfaces: {{From}} Microwaves to Visible},
  author = {Glybovski, Stanislav B. and Tretyakov, Sergei A. and Belov, Pavel A. and Kivshar, Yuri S. and Simovski, Constantin R.},
  date = {2016-05},
  journaltitle = {Physics Reports},
  volume = {634},
  pages = {1--72},
  publisher = {{North-Holland}},
  issn = {03701573},
  doi = {10.1016/j.physrep.2016.04.004}
}

@article{Guo2018,
  title = {High-Efficiency Visible Transmitting Polarizations Devices Based on the {{GaN}} Metasurface},
  author = {Guo, Zhongyi and Xu, Haisheng and Guo, Kai and Shen, Fei and Zhou, Hongping and Zhou, Qingfeng and Gao, Jun and Yin, Zhiping},
  date = {2018},
  journaltitle = {Nanomaterials},
  volume = {8},
  pages = {333},
  publisher = {{Multidisciplinary Digital Publishing Institute}},
  issn = {2079-4991},
  doi = {10.3390/nano8050333},
  number = {5}
}

@book{haykin2009neural,
  ids = {haykin2009neural},
  title = {Neural Networks and Learning Machines},
  author = {Haykin, Simon S.},
  date = {2009},
  edition = {3rd ed},
  publisher = {{Pearson Education}},
  location = {{New York}},
  isbn = {978-0-13-147139-9},
  pagetotal = {906}
}

@article{He2018,
  title = {High-Efficiency Metasurfaces: {{Principles}}, Realizations, and Applications},
  author = {He, Qiong and Sun, Shulin and Xiao, Shiyi and Zhou, Lei},
  date = {2018},
  journaltitle = {Advanced Optical Materials},
  volume = {6},
  pages = {1800415},
  publisher = {{John Wiley \& Sons, Ltd}},
  issn = {21951071},
  doi = {10.1002/adom.201800415},
  number = {19}
}

@article{Iadevaia6704,
  title = {Identification of {{Optimal Drug Combinations Targeting Cellular Networks}}: {{Integrating Phospho}}-{{Proteomics}} and {{Computational Network Analysis}}},
  shorttitle = {Identification of {{Optimal Drug Combinations Targeting Cellular Networks}}},
  author = {Iadevaia, S. and Lu, Y. and Morales, F. C. and Mills, G. B. and Ram, P. T.},
  date = {2010-09-01},
  journaltitle = {Cancer Research},
  shortjournal = {Cancer Research},
  volume = {70},
  pages = {6704--6714},
  publisher = {{American Association for Cancer Research}},
  issn = {0008-5472, 1538-7445},
  doi = {10.1158/0008-5472.CAN-10-0460},
  eprint = {https://cancerres.aacrjournals.org/content/70/17/6704.full.pdf},
  langid = {english},
  number = {17}
}

@article{Jahani2016,
  title = {All-Dielectric Metamaterials},
  author = {Jahani, Saman and Jacob, Zubin},
  date = {2016},
  journaltitle = {Nature Nanotechnology},
  volume = {11},
  pages = {23--36},
  publisher = {{Nature Publishing Group}},
  issn = {1748-3387},
  doi = {10.1038/nnano.2015.304},
  number = {1}
}

@book{Kennedy:2001:SI:370449,
  title = {Swarm Intelligence},
  author = {Kennedy, James F. and Eberhart, Russell C. and Shi, Yuhui},
  date = {2001},
  publisher = {{Morgan Kaufmann Publishers}},
  location = {{San Francisco}},
  isbn = {978-1-55860-595-4},
  pagetotal = {512},
  series = {The {{Morgan Kaufmann}} Series in Evolutionary Computation}
}

@article{Khorasaninejad2016,
  title = {Metalenses at Visible Wavelengths: {{Diffraction}}-Limited Focusing and Subwavelength Resolution Imaging},
  author = {Khorasaninejad, Mohammadreza and Chen, Wei Ting and Devlin, Robert C and Oh, Jaewon and Zhu, Alexander Y and Capasso, Federico},
  date = {2016},
  journaltitle = {Science},
  volume = {352},
  pages = {1190--1194},
  publisher = {{American Association for the Advancement of Science}},
  issn = {0036-8075},
  doi = {10.1126/science.aaf6644},
  eprint = {27257251},
  eprinttype = {pmid},
  number = {6290}
}

@article{Kuznetsov2016,
  title = {Optically Resonant Dielectric Nanostructures},
  author = {Kuznetsov, Arseniy I and Miroshnichenko, Andrey E and Brongersma, Mark L and Kivshar, Yuri S and Luk'yanchuk, Boris},
  date = {2016-11},
  journaltitle = {Science},
  volume = {354},
  pages = {aag2472},
  publisher = {{American Association for the Advancement of Science}},
  issn = {0036-8075},
  doi = {10.1126/science.aag2472},
  eprint = {27856851},
  eprinttype = {pmid},
  number = {6314}
}

@article{LESHNO1993861,
  title = {Multilayer Feedforward Networks with a Nonpolynomial Activation Function Can Approximate Any Function},
  author = {Leshno, Moshe and Lin, Vladimir Ya. and Pinkus, Allan and Schocken, Shimon},
  date = {1993-01},
  journaltitle = {Neural Networks},
  shortjournal = {Neural Networks},
  volume = {6},
  pages = {861--867},
  issn = {08936080},
  doi = {10.1016/S0893-6080(05)80131-5},
  langid = {english},
  number = {6}
}

@article{Li2019,
  title = {Efficient Polarization Beam Splitter Based on All-Dielectric Metasurface in Visible Region},
  author = {Li, Jing and Liu, Chang and Wu, Tiesheng and Liu, Yumin and Wang, Yu and Yu, Zhongyuan and Ye, Han and Yu, Li},
  date = {2019-12},
  journaltitle = {Nanoscale Research Letters},
  volume = {14},
  pages = {34},
  publisher = {{Springer US}},
  issn = {1931-7573},
  doi = {10.1186/s11671-019-2867-4},
  number = {1}
}

@article{Lin1004,
  title = {All-Optical Machine Learning Using Diffractive Deep Neural Networks},
  author = {Lin, Xing and Rivenson, Yair and Yardimci, Nezih T. and Veli, Muhammed and Luo, Yi and Jarrahi, Mona and Ozcan, Aydogan},
  date = {2018-09-07},
  journaltitle = {Science},
  shortjournal = {Science},
  volume = {361},
  pages = {1004--1008},
  publisher = {{American Association for the Advancement of Science}},
  issn = {0036-8075, 1095-9203},
  doi = {10.1126/science.aat8084},
  eprint = {https://science.sciencemag.org/content/361/6406/1004.full.pdf},
  langid = {english},
  number = {6406}
}

@article{Liu2018a,
  title = {Generative Model for the Inverse Design of Metasurfaces},
  author = {Liu, Zhaocheng and Zhu, Dayu and Rodrigues, Sean P. and Lee, Kyu-Tae and Cai, Wenshan},
  date = {2018-10},
  journaltitle = {Nano Letters},
  volume = {18},
  pages = {6570--6576},
  publisher = {{UTC}},
  issn = {1530-6984},
  doi = {10.1021/acs.nanolett.8b03171},
  number = {10}
}

@article{Ollanik2018a,
  title = {High-Efficiency All-Dielectric Huygens Metasurfaces from the Ultraviolet to the Infrared},
  author = {Ollanik, Adam J. and Smith, Jake A. and Belue, Mason J. and Escarra, Matthew D.},
  date = {2018},
  journaltitle = {ACS Photonics},
  volume = {5},
  pages = {1351--1358},
  publisher = {{American Chemical Society}},
  issn = {2330-4022},
  doi = {10.1021/acsphotonics.7b01368},
  number = {4}
}

@article{Overvig2019,
  title = {Dielectric Metasurfaces for Complete and Independent Control of the Optical Amplitude and Phase},
  author = {Overvig, Adam C. and Shrestha, Sajan and Malek, Stephanie C. and Lu, Ming and Stein, Aaron and Zheng, Changxi and Yu, Nanfang},
  date = {2019-12},
  journaltitle = {Light: Science \& Applications},
  volume = {8},
  pages = {92},
  publisher = {{Springer Science and Business Media LLC}},
  issn = {2047-7538},
  doi = {10.1038/s41377-019-0201-7},
  archivePrefix = {arXiv},
  arxivid = {1903.00578},
  eprint = {1903.00578},
  eprinttype = {arxiv},
  number = {1}
}

@article{resnet,
  title = {Deep {{Residual Learning}} for {{Image Recognition}}},
  author = {He, Kaiming and Zhang, Xiangyu and Ren, Shaoqing and Sun, Jian},
  date = {2015-12-10},
  url = {http://arxiv.org/abs/1512.03385},
  urldate = {2020-02-11},
  archivePrefix = {arXiv},
  eprint = {arXiv:1512.03385},
  eprinttype = {arxiv},
  primaryClass = {cs}
}

@article{Shrestha2018,
  title = {Broadband Achromatic Dielectric Metalenses},
  author = {Shrestha, Sajan and Overvig, Adam C. and Lu, Ming and Stein, Aaron and Yu, Nanfang},
  date = {2018-12},
  journaltitle = {Light: Science \& Applications},
  volume = {7},
  pages = {85},
  publisher = {{Nature Publishing Group}},
  issn = {2047-7538},
  doi = {10.1038/s41377-018-0078-x},
  number = {1}
}

@article{Slovick2017a,
  title = {Metasurface Polarization Splitter},
  author = {Slovick, Brian A. and Zhou, You and Yu, Zhi Gang and Kravchenko, Ivan I. and Briggs, Dayrl P. and Moitra, Parikshit and Krishnamurthy, Srini and Valentine, Jason},
  date = {2017},
  journaltitle = {Philosophical Transactions of the Royal Society A: Mathematical, Physical and Engineering Sciences},
  volume = {375},
  pages = {20160072},
  publisher = {{The Royal Society Publishing}},
  issn = {1364-503X},
  doi = {10.1098/rsta.2016.0072},
  number = {2090}
}

@article{Staude2017,
  title = {Metamaterial-Inspired Silicon Nanophotonics},
  author = {Staude, Isabelle and Schilling, Jörg},
  date = {2017},
  journaltitle = {Nature Photonics},
  volume = {11},
  pages = {274--284},
  publisher = {{Nature Publishing Group}},
  issn = {1749-4885},
  doi = {10.1038/nphoton.2017.39},
  number = {5}
}

@article{ToteroGongora2017,
  title = {Anapole Nanolasers for Mode-Locking and Ultrafast Pulse Generation},
  author = {Totero Gongora, Juan S. and Miroshnichenko, Andrey E. and Kivshar, Yuri S. and Fratalocchi, Andrea},
  date = {2017},
  journaltitle = {Nature Communications},
  volume = {8},
  pages = {15535},
  issn = {2041-1723},
  doi = {10.1038/ncomms15535},
  number = {1}
}

@article{Wang2016a,
  title = {Visible-Frequency Dielectric Metasurfaces for Multiwavelength Achromatic and Highly Dispersive Holograms},
  author = {Wang, Bo and Dong, Fengliang and Li, Qi-Tong and Yang, Dong and Sun, Chengwei and Chen, Jianjun and Song, Zhiwei and Xu, Lihua and Chu, Weiguo and Xiao, Yun-Feng and Gong, Qihuang and Li, Yan},
  date = {2016-08},
  journaltitle = {Nano Letters},
  volume = {16},
  pages = {5235--5240},
  publisher = {{American Chemical Society}},
  issn = {1530-6984},
  doi = {10.1021/acs.nanolett.6b02326},
  number = {8}
}

@article{Wang2018,
  title = {A Broadband Achromatic Metalens in the Visible},
  author = {Wang, Shuming and Wu, Pin Chieh and Su, Vin-Cent and Lai, Yi-Chieh and Chen, Mu-Ku and Kuo, Hsin Yu and Chen, Bo Han and Chen, Yu Han and Huang, Tzu-Ting and Wang, Jung-Hsi and Lin, Ray-Ming and Kuan, Chieh-Hsiung and Li, Tao and Wang, Zhenlin and Zhu, Shining and Tsai, Din Ping},
  date = {2018},
  journaltitle = {Nature Nanotechnology},
  volume = {13},
  pages = {227--232},
  publisher = {{Springer US}},
  issn = {1748-3387},
  doi = {10.1038/s41565-017-0052-4},
  number = {3}
}

@article{Wood2017,
  title = {All-Dielectric Color Filters Using {{SiGe}}-{{Based}} Mie Resonator Arrays},
  author = {Wood, Thomas and Naffouti, Meher and Berthelot, Johann and David, Thomas and Claude, Jean-Benoît and Métayer, Léo and Delobbe, Anne and Favre, Luc and Ronda, Antoine and Berbezier, Isabelle and Bonod, Nicolas and Abbarchi, Marco},
  date = {2017-04},
  journaltitle = {ACS Photonics},
  volume = {4},
  pages = {873--883},
  publisher = {{American Chemical Society}},
  issn = {2330-4022},
  doi = {10.1021/acsphotonics.6b00944},
  number = {4}
}

@article{Yang2017,
  title = {Dielectric Nanoresonators for Light Manipulation},
  author = {Yang, Zhong-Jian and Jiang, Ruibin and Zhuo, Xiaolu and Xie, Ya-Ming and Wang, Jianfang and Lin, Hai-Qing},
  date = {2017-07},
  journaltitle = {Physics Reports},
  volume = {701},
  pages = {1--50},
  publisher = {{North-Holland}},
  issn = {03701573},
  doi = {10.1016/j.physrep.2017.07.006}
}

@article{Yu2014,
  title = {Flat Optics with Designer Metasurfaces},
  author = {Yu, Nanfang and Capasso, Federico},
  date = {2014-02},
  journaltitle = {Nature Materials},
  volume = {13},
  pages = {139--150},
  publisher = {{Nature Publishing Group}},
  issn = {1476-1122},
  doi = {10.1038/nmat3839},
  number = {2}
}

@article{Yu2015,
  title = {High-Transmission Dielectric Metasurface with 2{{$\pi$}} Phase Control at Visible Wavelengths},
  author = {Yu, Ye Feng and Zhu, Alexander Y and Paniagua-Domínguez, Ramón and Fu, Yuan Hsing and Luk'yanchuk, Boris and Kuznetsov, Arseniy I},
  date = {2015-07},
  journaltitle = {Laser \& Photonics Reviews},
  volume = {9},
  pages = {412--418},
  issn = {18638880},
  doi = {10.1002/lpor.201500041},
  number = {4}
}

@article{Zhou2017,
  title = {Efficient Silicon Metasurfaces for Visible Light},
  author = {Zhou, Zhenpeng and Li, Juntao and Su, Rongbin and Yao, Beimeng and Fang, Hanlin and Li, Kezheng and Zhou, Lidan and Liu, Jin and Stellinga, Daan and Reardon, Christopher P. and Krauss, Thomas F. and Wang, Xuehua},
  date = {2017},
  journaltitle = {ACS Photonics},
  volume = {4},
  publisher = {{American Chemical Society}},
  issn = {2330-4022},
  doi = {10.1021/acsphotonics.6b00740},
  number = {3}
}

@article{Ren2019,
  title = {Metasurface Orbital Angular Momentum Holography},
  author = {Ren, Haoran and Briere, Gauthier and Fang, Xinyuan and Ni, Peinan and Sawant, Rajath and H\'eron, S\'ebastien and Chenot, S\'ebastien and V\'ezian, St\'ephane and Damilano, Benjamin and Br\"andli, Virginie and Maier, Stefan A. and Genevet, Patrice},
  date = {2019-07-19},
  journaltitle = {Nature Communications},
  shortjournal = {Nat Commun},
  volume = {10},
  pages = {1--8},
  issn = {2041-1723},
  doi = {10.1038/s41467-019-11030-1},
  langid = {english},
  number = {1}
}

@article{Shibanuma2016,
  title = {Unidirectional Light Scattering with High Efficiency at Optical Frequencies Based on Low-Loss Dielectric Nanoantennas},
  author = {Shibanuma, Toshihiko and Albella, Pablo and Maier, Stefan A.},
  date = {2016-07-21},
  journaltitle = {Nanoscale},
  shortjournal = {Nanoscale},
  volume = {8},
  pages = {14184--14192},
  issn = {2040-3372},
  doi = {10.1039/C6NR04335F},
  langid = {english},
  number = {29}
}

@article{Huang2018,
  title = {Metasurface Holography: From Fundamentals to Applications},
  shorttitle = {Metasurface Holography},
  author = {Huang, Lingling and Zhang, Shuang and Zentgraf, Thomas},
  date = {2018},
  journaltitle = {Nanophotonics},
  volume = {7},
  pages = {1169--1190},
  doi = {10.1515/nanoph-2017-0118},
  number = {6}
}

@article{Zheng2017,
  title = {Dual Field-of-View Step-Zoom Metalens},
  author = {Zheng, Guoxing and Wu, Weibiao and Li, Zile and Zhang, Shuang and Mehmood, Muhammad Qasim and He, Ping'an and Li, Song},
  date = {2017-04-01},
  journaltitle = {Optics Letters},
  shortjournal = {Opt. Lett., OL},
  volume = {42},
  pages = {1261--1264},
  issn = {1539-4794},
  doi = {10.1364/OL.42.001261},
  langid = {english},
  number = {7}
}

@Article{Guo2017,
  author       = {Guo, Zhongyi and Zhu, Lie and Shen, Fei and Zhou, Hongping and Gao, Rongke},
  date         = {2017-01-30},
  journaltitle = {RSC Advances},
  title        = {Dielectric Metasurface Based High-Efficiency Polarization Splitters},
  doi          = {10.1039/C6RA27741A},
  issn         = {2046-2069},
  number       = {16},
  pages        = {9872--9879},
  volume       = {7},
  langid       = {english},
  publisher    = {{The Royal Society of Chemistry}},
  shortjournal = {RSC Adv.},
}

@Article{Chen2019,
  author       = {Chen, Wei Ting and Zhu, Alexander Y. and Sisler, Jared and Bharwani, Zameer and Capasso, Federico},
  date         = {2019-12},
  journaltitle = {Nature Communications},
  title        = {A Broadband Achromatic Polarization-Insensitive Metalens Consisting of Anisotropic Nanostructures},
  doi          = {10.1038/s41467-019-08305-y},
  issn         = {2041-1723},
  number       = {1},
  pages        = {355},
  volume       = {10},
}


\clearpage

\begin{figure*} \centering \includegraphics[width=\textwidth]{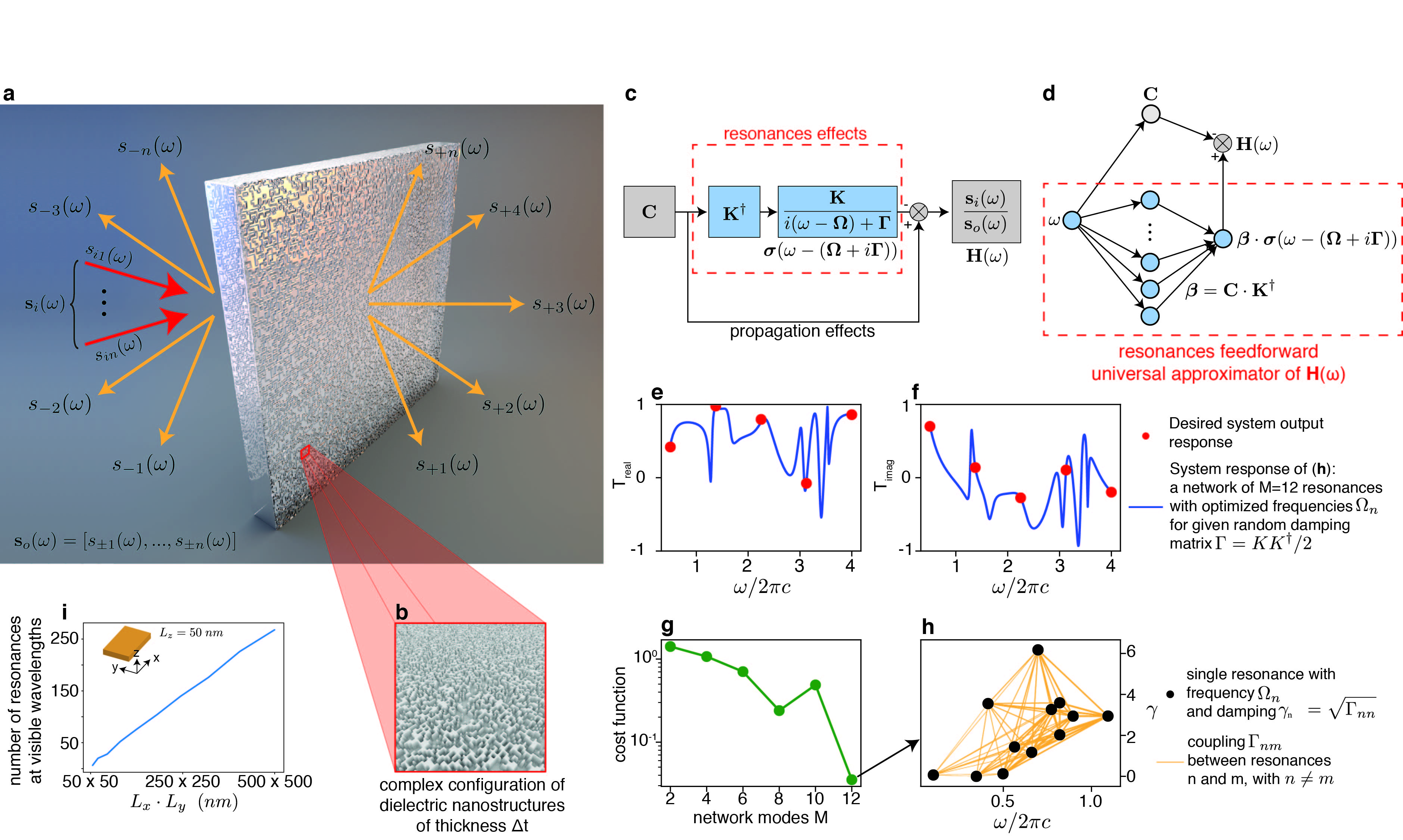}
\caption{ \label{funo} \textbf{Universal flat-optics approximators: general idea.} (a) Problem setup composed by a flat-optical surface made by resonant nanostructures (b) with input $\mb s_{i}$ and output $\mb s_{o}$ scattered waves. (i) Dependency of the number of resonances on the dimensions of a nanostructure. (c) Block-diagram of the input-output transfer function $\mb H(\omega)=\frac{\mb s_{i}(\omega)}{\mb s_{o}(\omega)}$. (d) Equivalent representation of (c) with a feedforward single hidden layer neural network modeling the effects of the resonances. (e-h) Demonstration of the  universal representation behavior of (d): an arbitrarily defined system response (e-f, red markers) is obtained by tuning the resonances $\boldsymbol{\Omega}$ of the network for given initial weights $\boldsymbol{\beta}$, couplings $\mb K$ and damping $\boldsymbol{\Gamma}$. The problem is solved by minimizing a cost function (g) by using an increasing number of resonances $M$, which defines the size of $\boldsymbol{\Omega}$. (h) Network configuration that represents the desired response (e-f, solid line). The general demonstration of this result for an arbitrary structure is carried out in Suppl. Note II.  
} \end{figure*}

\clearpage

\begin{figure*} \centering
    \includegraphics[width=.99\textwidth]{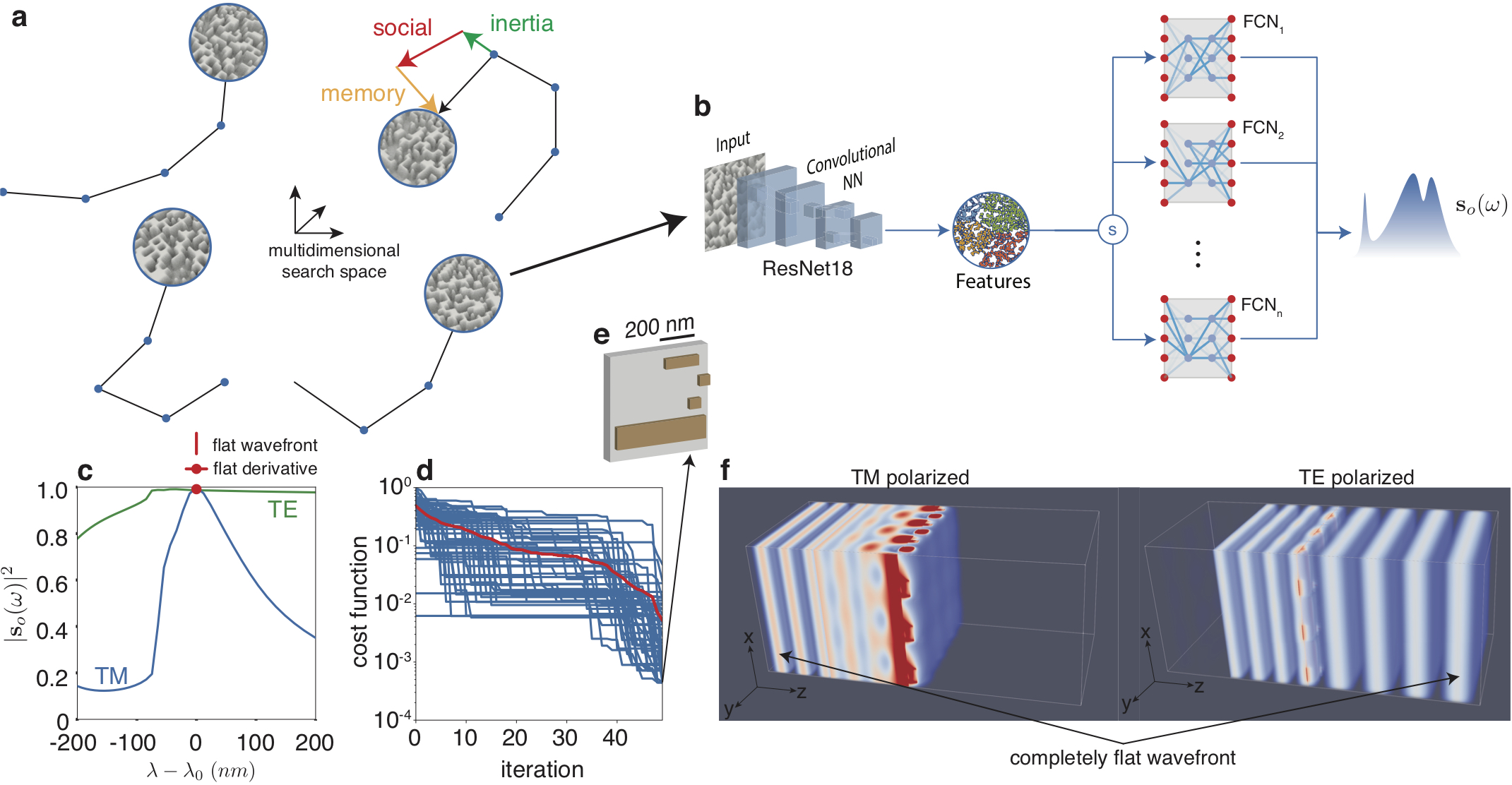} \caption{
	\label{fdue} \textbf{ALFRED design idea and example of application.} (a) Swarm parallel optimizer that searches the multidimensional space of solutions with an array of tentative geometries that use a cooperative scheme based on self and social interactions. (b) Predictor on each tentative solution: a convolutional neural network (CNN) extracts features from an input geometry and predicts the spectral response $\mb s_o(\omega)$ through a series of fully connected networks trained on different material thicknesses that are automatically selected by the switch (S). (c-f) Design example of polarizer beam splitter. (c) Target response that maximizes the normal transmission for TE (green) and normal reflection for TM (blue) polarizations with flat first order derivative at the target design wavelength $\lambda_0=900$~nm. (d) Cost function for progressive iterations of ALFRED (blue: particles, red: mean value). (e) Designed structure with gray unit cell and brown nanoresonators. (f) FDTD simulations of (e) for TE and TM polarizations at $\lambda=\lambda_0$.}  \end{figure*}

\clearpage

\begin{figure*} \centering
    \includegraphics[width=.99\textwidth]{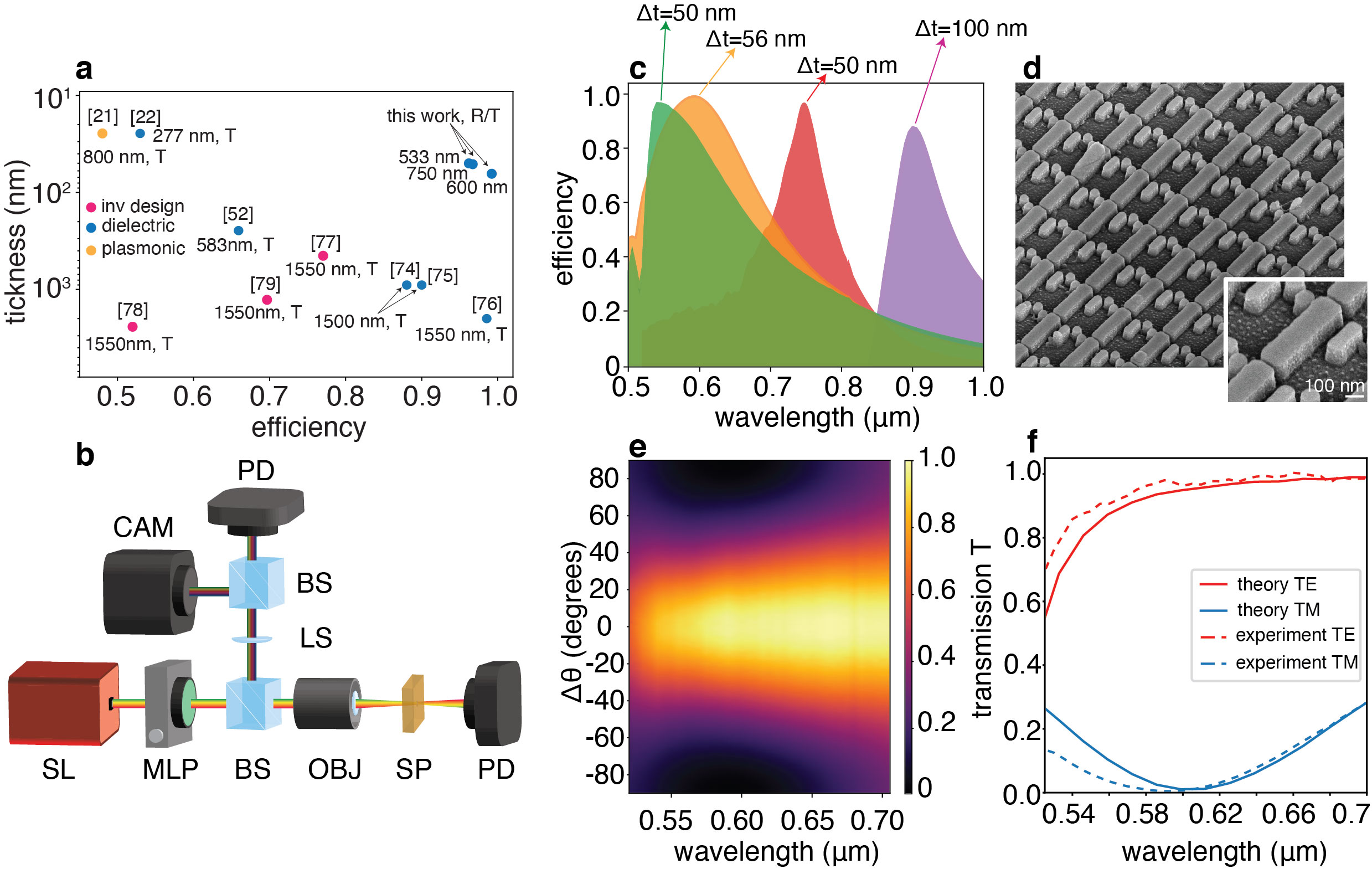} \caption{
    \label{Polarizer_characterisation}
	\textbf{Experimental results on polarizing beam splitters.} (a) Flat-optics state of the art at different wavelengths for dielectric and metallic systems (R: reflection, T: transmission, R/T: both). (b) Experimental setup used to chracterize the samples (SL: supercontinuum laser, MLP: motorized linear polarizer, BS: beam splitter, LS: lens, OBJ: objective, CAM: CMOS camera, PD: calibrated photodetector, SP: sample). (c) experimentally measured efficiencies at different wavelengths for four different samples working around the common laserline wavelengths $533$~nm, $600$~nm, $750$~nm, and $900$~nm. (d) SEM image of a characteristic sample for the polarizer designed at $900$~nm. (e) Transmission efficiency of the $600$~nm polarizer for different wavelengths and input polarization angles $\Delta\theta$. (f) Comparison between theory (solid line) and experiment (dashed line) for $600$~nm polarizer transmission efficiency. \nocite{Pelletier2006,Hong2007b,Li2019,Guo2017b,Guo2017,Wang2007d,Shen2014,Shen2015,Frandsen2016}} \end{figure*}

\clearpage

\begin{figure*} \centering
    \includegraphics[width=0.8\textwidth]{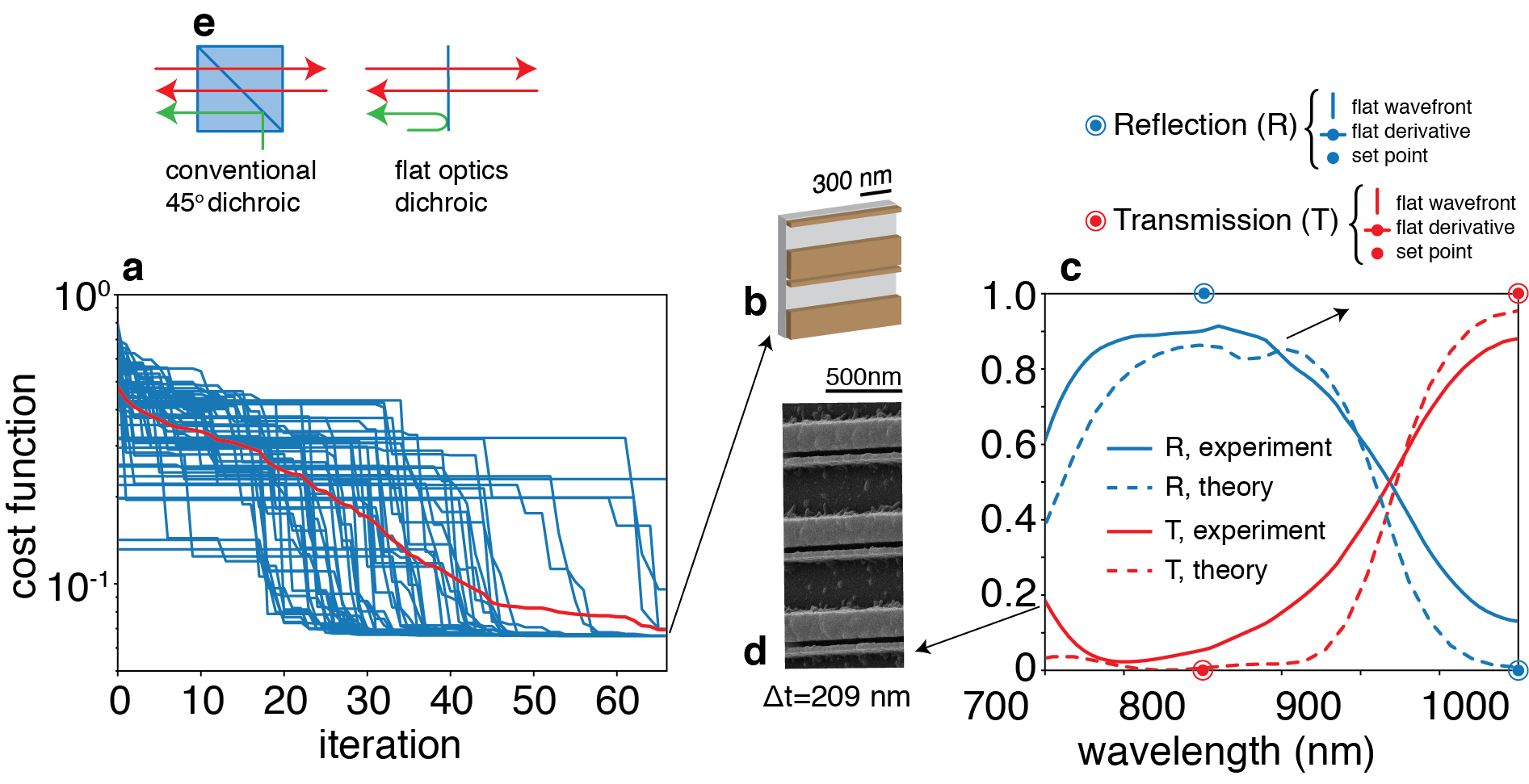}
    \caption{\textbf{Flat-optics dichoric mirror.} (a) Cost function of the target design for increasing iterations of ALFRED (blue: particles, red: mean value) and (b) sample design found. The dichroic component is designed by $4$ target points (c, blue/red markers), each characterized by a point with flat derivative and flat wavefront. Panel (c) compares FDTD theoretical (dashed line) and experimental (solid line) results. (d) SEM image of a fabricated sample. (e) Conventional dichroic geometry, requiring $45^\circ$ impinging light, versus flat optics structure working at normal incidence.}
\label{Dichroic_mirror} \end{figure*}

\clearpage

\begin{figure*} \centering \includegraphics[width=\textwidth]{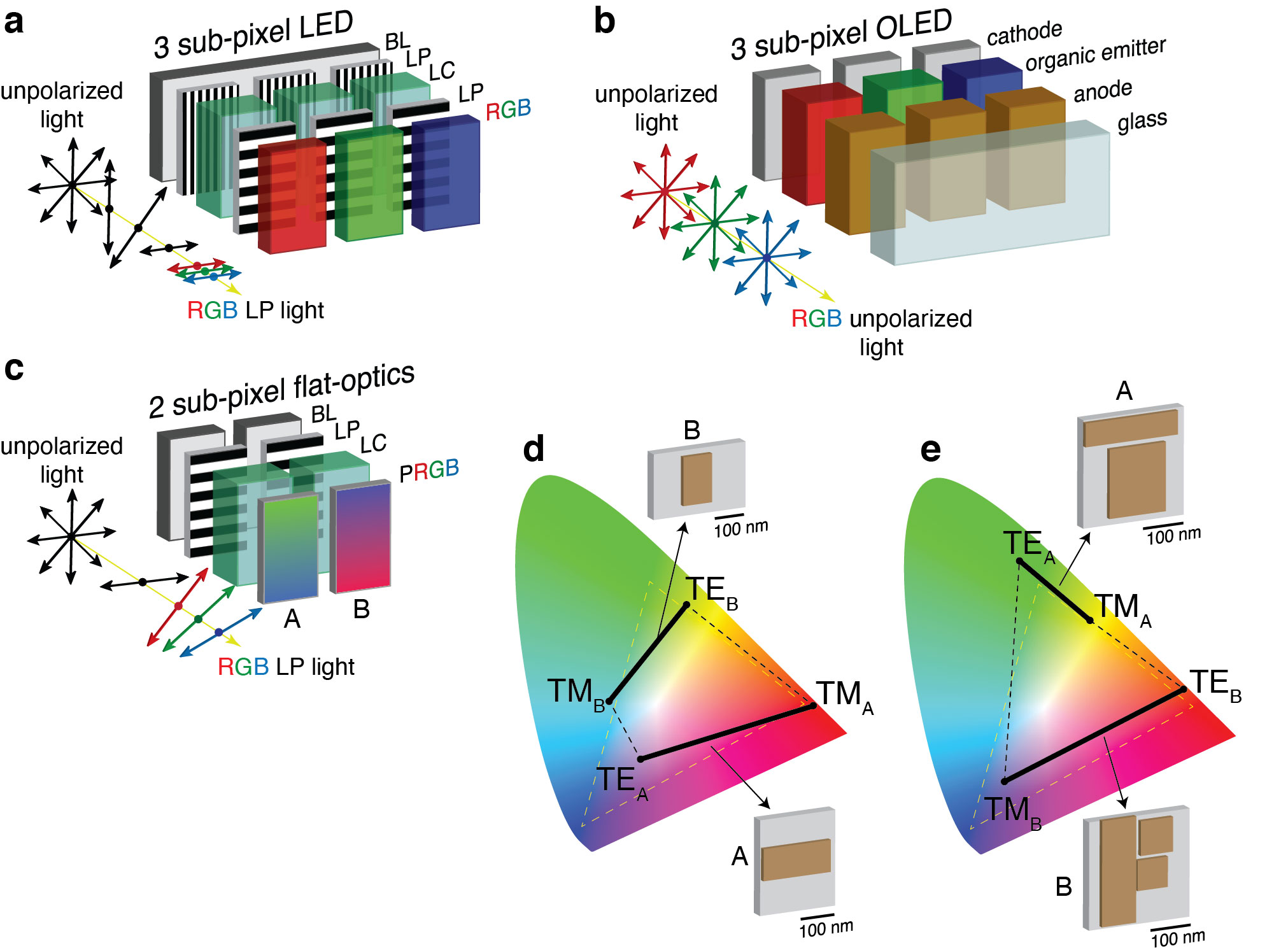}
    \caption{\textbf{Two sub-pixel flat optics color display: idea and design.} (a-b) Basic working principle of current LCD/OLED color technologies based on three red, green and blue (RGB) sub-pixel color transmission units. (c) Proposed flat-optics technology based on two polarization controlled sub-pixels A and B. (d-e) ALFRED designs of A and B (3D brown boxes with gray unit cell) with chromaticity gamuts contained in (d) and exceeding (e) the standard RGB colorspace (dashed yellow line). In panels (d)-(e), the chromaticity obtained when the input light varies between $TE_x$ and $TM_x$ polarizations for samples $A,B$ is indicated as a solid black thick line, while the total gamut obtained by mixing the two sub-pixels is the four point rhomboid delimited by the dashed black lines.}\label{pixel2-figure}
\end{figure*}

\begin{figure*} \centering
    \includegraphics[width=\textwidth]{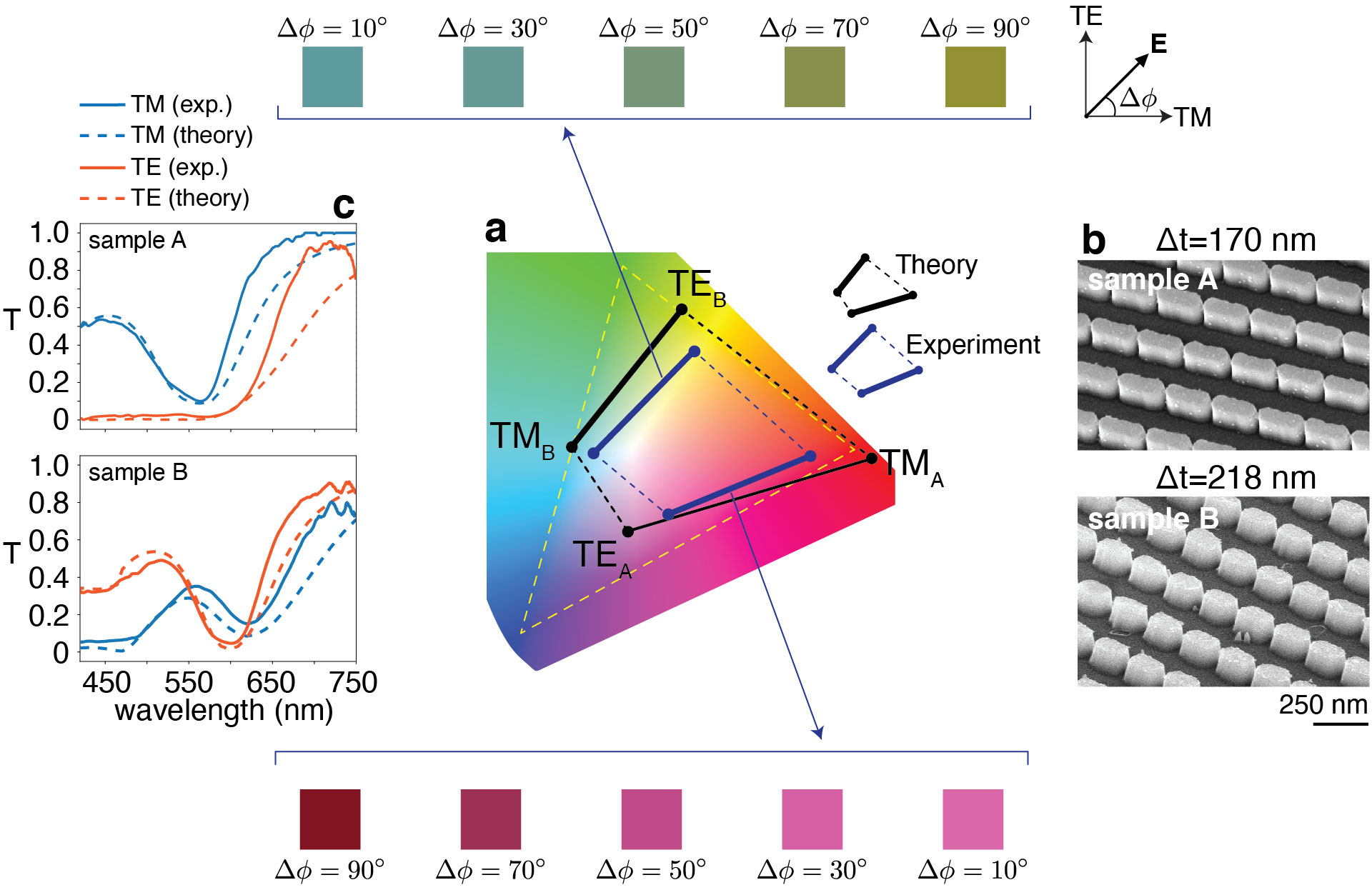}
    \caption{\textbf{Characterisation of the fabricated two sub-pixel color samples.} (a) Theoretical (black rhomboid) and experimental (blue rhomboid) chromaticity gamut of samples A and B illustrated in the SEM of panel (b). The corresponding color for different input polarization angles $\Delta\phi$ are reported in panel (a). (c) FDTD calculated theoretical transmission (T) response (dashed line) compared to experimental measurements (solid line).  }  \label{pixel-figure}
\end{figure*}

\end{document}


\title{Broadband vectorial ultra-flat optics with experimental efficiency up to 99\% in the visible via universal approximators \\
\vspace{1em} SI}


\author{F. Getman, M. Makarenko, A. Burguete-Lopez \&
A. Fratalocchi}

\maketitle




\clearpage

\section*{Supplementary Note I}
The activation function $\boldsymbol\sigma=\frac{\mb K}{i(\omega-\mb \Omega)+\mb\Gamma}$ is obtained from the solution of the following differential equation:
\begin{equation}
    \label{sig0}
    \frac{d\boldsymbol{\tilde\sigma}}{dt}=\left(i\boldsymbol \Omega-\boldsymbol\Gamma\right)\boldsymbol{\tilde\sigma}+\mb K \cdot \delta(t),
\end{equation}
with $\bs=\int d\omega\boldsymbol\sigma(\omega) e^{i\omega t}$ being the inverse Fourier transform of $\boldsymbol{\sigma}$ and $\delta(t)$ the diract delta function, as the reader can verify by direct substitution. The solution of Eq. \eqref{sig0} reads:
\begin{equation}
\label{exp01}
\bs(t)=e^{\mb At}\cdot(\bs_0+\mb K),
\end{equation}
with $\bs_0=\bs(t=0)$ the initial state and $e^{\mb At}$ being the matrix exponential, with $\mb A=i\boldsymbol\Omega-\boldsymbol\Gamma$. The matrix exponential is expressed in closed form by diagonalizing $\mb A=\mb Q\boldsymbol\Lambda \mb Q^{-1}$, with $\Lambda_{mm'}=\lambda_m\delta_{mm'}$ being a diagonal matrix of complex eigenvalues $\lambda_m=i\alpha_m-\beta_m$:
\begin{equation}
\label{exp00}
e^{\mb At}=\mb Qe^{\mb \Lambda t}\mb Q^{-1}.
\end{equation}
In Eq. \eqref{exp00}, $\mb Q=\mb Q(\boldsymbol\Omega,\boldsymbol \Gamma)$ and $\boldsymbol\Lambda=\boldsymbol\Lambda(\boldsymbol\Omega,\boldsymbol\Gamma)$ originate from the eigenvectors and eigenvalues of $\mb A=i\boldsymbol\Omega-\boldsymbol\Gamma$ and as such they are functions of $\boldsymbol{\Omega}$ and $\boldsymbol{\Gamma}$. The elements $\tilde\sigma_{mn}$ are calculated by substituting \eqref{exp00} into \eqref{exp01}:
\begin{align}
\label{at0}
    \tilde\sigma_{mn}(t)=\sum_{h}\mu_{mnh}e^{(i\alpha_{h}-\beta_{ h})t},
\end{align}
and expressed as the sum of complex damped exponentials with $\mu_{mnh}=\mu_{mnh}(\boldsymbol\Omega,\boldsymbol\Gamma)$ coefficients arising from the matrix product of $e^{\mb At}\cdot(\bs_0+\mb K)$. By applying the Fourier transform on Eq. \eqref{at0}, we obtain the expression of the elements $\sigma_{mn}$ of $\boldsymbol{\sigma}$:
\begin{equation}
\label{at1}
     \sigma_{mn}(\omega)=\frac{1}{\sqrt{2\pi}}\sum_{h}\frac{\mu_{mnh}}{i(\omega-\omega_{h})+\gamma_{h}},
\end{equation}
represented as a complex rational function with poles $\omega_h+i\gamma_h$, and coefficients $\mu_{mnh}$.

\section*{Supplementary Note II}
We considered a general rational function:
\begin{equation}
    \label{eq0}
    y(\omega)=\frac{\sum_{n=0}^{N_\alpha}\alpha_n\cdot \omega^n}{\sum_{n=0}^{N_\beta}\beta_n\cdot \omega^n},
\end{equation}
with $N_\alpha+1$ and $N_\beta+1$ coefficients $\alpha_n$ and $\beta_n$, respectively. Without loss of generality and for the sake of simplicity, we assumed that the function $y_{out}$ is real. If we control $N_\alpha+N_\beta+2$ coefficients $\alpha_n$ and $\beta_n$, we can then can uniquely determine the behavior of the output function $y_{out}(\omega_m)$ at a set of $m=1,...,M=N_\alpha+N_\beta+1$ spectral frequencies $\omega_m$. To demonstrate this result, we impose the conditions $y_{out}(\omega_m)=y_{m}(\omega_m)$ on Eq. \eqref{eq0} and obtain the following algebraic system:
\begin{equation}
    \label{lin0}
    \begin{bmatrix}
    \mathrm{diag}\left(y(\omega_1),...,y(\omega_M)\right)\cdot V_{N_\beta+1} &V_{N_\alpha+1}
    \end{bmatrix}
    \begin{bmatrix}
    \boldsymbol\beta\\\
    -\boldsymbol\alpha
    \end{bmatrix}=0,
\end{equation}
with diag being the diagonal matrix for the $M$ function values $y(\omega_m)$ at the target points, $V_i$ the first $i$ columns of the Vandermonde matrix $V$ generated at the sample points $\omega_m$ with components $V_{ij}=\omega_i^{j-1}$, $\boldsymbol\alpha=[\alpha_0,...,\alpha_{N_\alpha}]$ and $\boldsymbol\beta=[\beta_0,...,\beta_{N_\beta}]$ being the rational coefficients sought. The solution to Eq. \eqref{eq0} requires one to find a null vector of the $M\times (N_\alpha+N_\beta+2)$ matrix:
\begin{equation}
    \label{ns0}
    \begin{bmatrix}
    \mathrm{diag}\left(y(\omega_1),...,y(\omega_M)\right)\cdot V_{N_\beta+1} &V_{N_\alpha+1}
    \end{bmatrix}.
\end{equation}
For $M=N_\alpha+N_\beta+1$, the matrix \eqref{ns0} has a null space of dimension $N_\alpha+N_\beta+2-M=1$, and each null vector gives an exact solution in \eqref{eq0}~\cite{10.5555/17318}. For $M<N_\alpha+N_\beta+1$, the system is under determined and the null space of \eqref{ns0} has dimension $N_\alpha+N_\beta+2-M>1$. In this case, any null vector among the space of $\infty^{N_\alpha+N_\beta+2-M}$ provides an exact solution to the problem.\\
For $M>N_\alpha+N_\beta+1$, the system is over determined and the solution to Eqs. \eqref{eq0} can be found in the least square sense by constrained minimization:
\begin{widetext}
\begin{align}
    \label{lsq0}
    &\mathrm{Minimize}\left\|\begin{bmatrix}
    \mathrm{diag}\left(y(\omega_1),...,y(\omega_M)\right)\cdot V_{N_\beta+1} &V_{N_\alpha+1}
    \end{bmatrix}
    \begin{bmatrix}
    \boldsymbol\beta\\
    -\boldsymbol\alpha
    \end{bmatrix}\right\|, &\mathrm{with}\;\;\left\|\boldsymbol\alpha\right\|+\left\|\boldsymbol\beta\right\|=1,
\end{align}
\end{widetext}
with $\|.\|$ being the vector norm. The constraint $\left\|\boldsymbol\alpha\right\|+\left\|\boldsymbol\beta\right\|=1$ is introduced to avoid the trivial solution $\boldsymbol\beta=\boldsymbol\alpha=0$.\\
In the case of Eq. (1) of the main text, the transfer function $H$ is written as follows:
\begin{equation}
    H(\omega)=\frac{s_{out}}{s_{in}}=\left[...,\frac{s_{-n}}{s_{in}},...,\frac{s_{+n}}{s_{in}}\right]=C(\omega)-\beta\sigma.
\end{equation}
We begin by considering the case $C(\omega)=C$ independent of the frequency $\omega$. In this case, Eq. \eqref{at1} shows that $H$ is a complex rational function with coefficients depending on the resonance frequencies $\Omega$ and mode dampings $\Gamma$. The control of $2M$ resonances $\Omega_n$ implies the control of $2M$ coefficients in the transfer function $H$. Equation \eqref{lin0} shows that we can therefore control the system response in amplitude and phase exactly at $M-1$ spectral points in both real and imaginary part, or equivalently, amplitude and phase in a single scattering channel. At any number of points $\ge M$, the problem can be solved in the least square sense via \eqref{lsq0}.\\
This result applies also in the case where $C(\omega)$ is frequency dependent. In this case every condition $H(\omega_l)=H_l$ generates a rational function with different constants $C_l=C(\omega_l)$, maintaining the validity of Eqs. \eqref{lin0}-\eqref{lsq0}.

\section*{Supplementary Note III}
We considered an optical network with $N=10$ output scattering channels, initialized with random scattering $C=e^{iH}$, with $H$ being a random symmetric matrix with elements drawn from a uniform distribution between zero and $2\pi$. The random couplings $K=A+iB$ are initialized with random matrices $A$ and $B$, each with elements drawn from a uniform distribution. We then initialized the dampings to $\Gamma=\frac{KK^\dag}{2}$. The transmission $T$ is measured in the first channel $s_{+1}$ computed at $n=1$.   

\bibliography{metasurfaces_article_biblatex}


\begin{figure*}
       \centering
	\includegraphics[width=\textwidth]{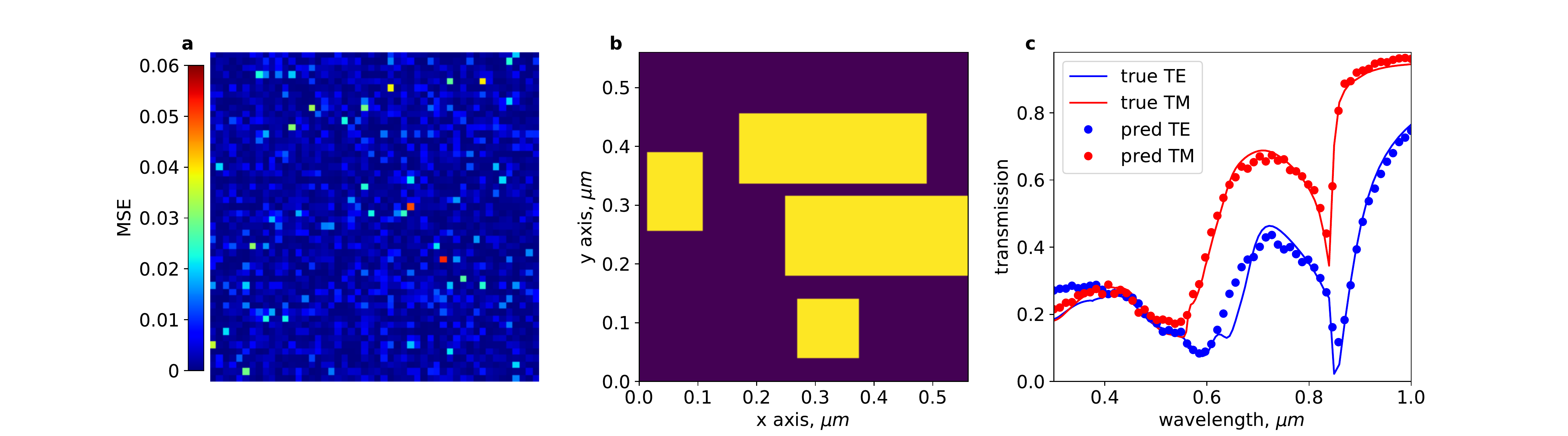}
	\renewcommand{\figurename}{Supplementary Figure}
	\caption{
		\label{idea}	
					ALFRED performance on 2500 randomly chosen samples from a test dataset \textbf{a}. The colors represent the mean squared error between the ground truth and the predicted spectra for each sample. An example of the output from the predictor part of  ALFRED for the random test sample  \textbf{c}. The blue and red curves on the figure represent TE and TM spectra respectively, for the true values and dotted lines represents neural network predictions. Figure \textbf{b} represents a masked array structure for the given sample.} 
\end{figure*}

\begin{figure*}
       \centering
	\includegraphics[width=0.5\textwidth]{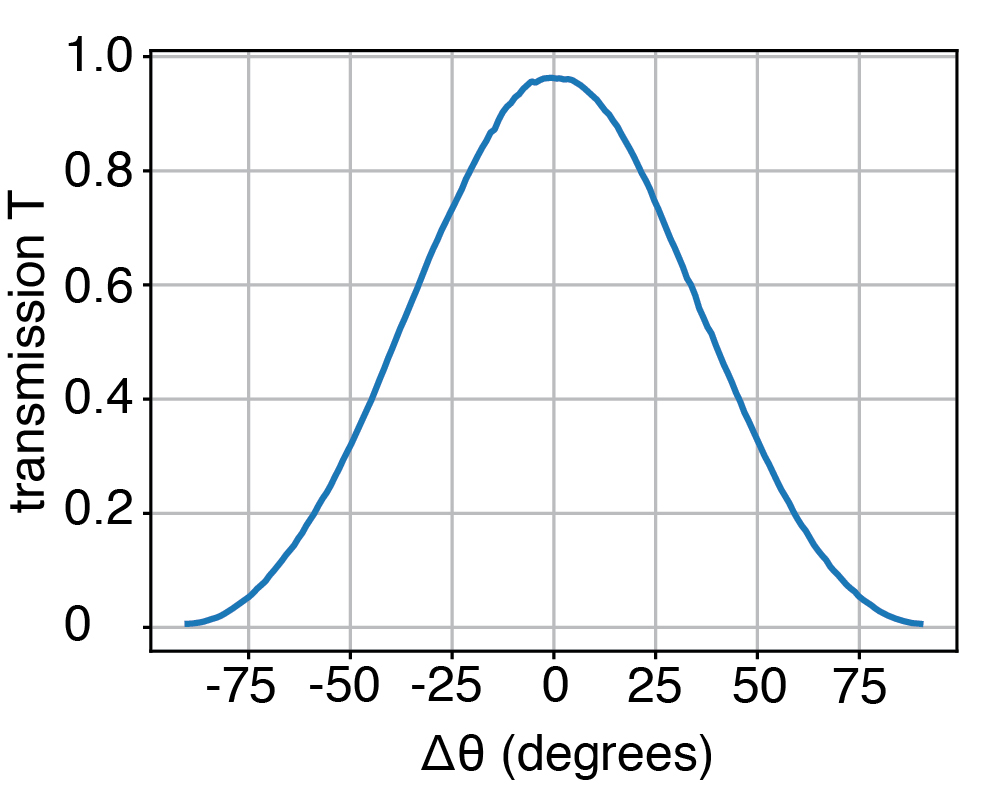}
	\renewcommand{\figurename}{Supplementary Figure}
	\caption{
		\label{idea}	
					Transmission efficiency of $600$~nm polarizer at the designed wavelength of $600$~nm wavelength versus input polarization angles $\Delta\theta$.} 
\end{figure*}